\documentclass[runningheads]{llncs}

\usepackage{eccv}

\usepackage{eccvabbrv}

\usepackage{booktabs}

\usepackage[accsupp]{axessibility}  
\usepackage{footnote}	
\usepackage{latexsym}
\usepackage{graphicx}
\usepackage{multicol,multirow}
\usepackage{amsmath,amssymb,amsfonts}
\usepackage{mathrsfs}
\usepackage{rotating}
\usepackage{appendix}
\usepackage{ifpdf}
\usepackage[T1]{fontenc}
\usepackage{times}
\usepackage{newtxmath}
\usepackage{textcomp}%
\usepackage{xcolor}%
\usepackage{hyperref}
\usepackage{lipsum}
\usepackage{float,placeins}
\usepackage{marvosym}
\usepackage[normalem]{ulem}
\hypersetup{
    colorlinks=true,
    linkcolor=blue,
    filecolor=magenta,      
    urlcolor=cyan,
    pdftitle={Overleaf Example},
    pdfpagemode=FullScreen,
    }

\usepackage{orcidlink}

\begin{document}

\title{The Quest for Early Detection of Retinal Disease: 3D CycleGAN-based Translation of Optical Coherence Tomography into Confocal Microscopy} 

\titlerunning{3D CycleGAN for OCT-Confocal Image Translation}

\author{Xin Tian\inst{1}\orcidlink{0000-0003-1168-5298}$^{(\textrm{\Letter})}$ \and
Nantheera Anantrasirichai\inst{1}\orcidlink{0000-0002-2122-5781} \and
Lindsay Nicholson\inst{2}\orcidlink{0000-0002-6942-208X} \and
Alin Achim\inst{1}\orcidlink{0000-0002-0982-7798}
}

\authorrunning{X. Tian et al.}
%
%
\institute{Visual Information Laboratory, University of Bristol, Bristol, UK \and Autoimmune Inflammation Research, University of Bristol, Bristol, UK\\
\email{\{xin.tian, n.anantrasirichai, l.nicholson, alin.achim\}@bristol.ac.uk}}

\maketitle

\begin{abstract}
Optical coherence tomography (OCT) and confocal microscopy are pivotal in retinal imaging, offering distinct advantages and limitations. In vivo OCT offers rapid, non-invasive imaging but can suffer from clarity issues and motion artifacts, while ex vivo confocal microscopy, providing high-resolution, cellular-detailed color images, is invasive and raises ethical concerns. To bridge the benefits of both modalities, we propose a novel framework based on unsupervised 3D CycleGAN for translating unpaired in vivo OCT to ex vivo confocal microscopy images. This marks the first attempt to exploit the inherent 3D information of OCT and translate it into the rich, detailed color domain of confocal microscopy. We also introduce a unique dataset, OCT2Confocal, comprising mouse OCT and confocal retinal images, facilitating the development of and establishing a benchmark for cross-modal image translation research. Our model has been evaluated both quantitatively and qualitatively, achieving Fréchet Inception Distance (FID) scores of 0.766 and Kernel Inception Distance (KID) scores as low as 0.153, and leading subjective Mean Opinion Scores (MOS). Our model demonstrated superior image fidelity and quality with limited data over existing methods. Our approach effectively synthesizes color information from 3D confocal images, closely approximating target outcomes and suggesting enhanced potential for diagnostic and monitoring applications in ophthalmology.

\end{abstract}

\section{Introduction}
\label{sec:intro}
Multimodal retinal imaging is critical in ophthalmological evaluation, enabling comprehensive visualization of retinal structures through imaging techniques such as fundus photography, Optical Coherence Tomography (OCT), fluorescein angiography (FFA), and confocal microscopy~\cite{retinareview2010, kang2021multimodal, meleppat2021vivo}. Each imaging modality manifests different characteristics of retinal structure, such as blood vessels, retinal layers and cellular distribution. Thus, integrating images from these techniques can help with tasks such as retinal segmentation~\cite{morano2020multimodal,hu2012multimodal}, image-to-image translation (I2I)~\cite{vidal2023image, abdelmotaal2022bridging}, and image fusion~\cite{el2024multimodality, tian2019multimodal}. Thereby improving the diagnosis and treatment of a wide range of diseases, from Diabetic Retinopathy (DR), macular degeneration, to glaucoma~\cite{kang2021multimodal, el2024multimodality}.

Among these retinal imaging modalities, confocal microscopy and OCT stand as preeminent methodologies for three-dimensional retinal imaging, each offering unique insights into the complexities of retinal anatomy. Confocal microscopy is a powerful ophthalmic imaging technique that generates detailed, three-dimensional images of biological tissues. It utilizes point illumination and point detection to visualize specific cells or structures. This allows for exceptional depth discrimination and detailed structural analysis. This technique is particularly adept at revealing the intricate cellular details of the retina, crucial for the detection of abnormalities or pathologies~\cite{paula2010microstructure,invivoconfocal}. Although in vivo confocal microscopy enables non-invasive examination of the ocular surface, its application is confined to imaging superficial retinal layers and is constrained by a small field of view, as well as by the impact of normal microsaccadic eye movements on the quality of the images~\cite{invivoconfocal2, al2010ex, bhattacharya2022segmentation}. On the other hand, ex vivo confocal microscopy, requiring tissue removal from an organism, is invaluable in research settings as it offers enhanced resolution, no movement artifacts, deeper and detailed structural information, in vitro labelling of specific cell markers and visualisation of cellular-level pathology which is not achievable with in vivo methods~\cite{Microvasculaturconfocal, ArteryVeinconfocal}. 
The high-resolution capabilities of confocal microscopy allow for a more granular assessment of tissue health. This includes clearer visualization of changes in the appearance and organization of Retinal Pigment Epithelium (RPE) cells, crucial in the pathogenesis of AMD. Moreover, it is vital for observing microvascular changes such as microaneurysms and capillary dropout in diabetic retinopathy, and for monitoring neovascularization and its response to treatment, offering detailed insights into therapeutic effectiveness. These attributes make ex vivo confocal microscopy an essential tool for comprehensive retinal research. While ex vivo confocal microscopy can only be used to image human retina post-mortem, making it ineligible for use in regular clinical screening. Thus, it is notably beneficial to use murine retinal studies as the mouse retina shares significant anatomical and physiological similarities with the human retina~\cite{humanconfocal, mouseconfocal}. However, ex vivo confocal imaging requires tissue removal with the potential to introduce artefacts through extracting and flattening the retina. Furthermore, the staining process can lead to over-coloring, uneven color distribution, or incorrect coloring, potentially complicating the interpretation of pathological features. 

The OCT, on the other hand, is a non-invasive (in vivo) tomographic imaging technique that provides three-dimensional images of the retinal layers, offering a comprehensive view of retinal anatomy. It boasts numerous advantages, such as rapid acquisition times and the ability to provide detailed cross-sectional grayscale images, which yield structural information at the micrometer scale. Clinically, OCT is utilized extensively for its objective and quantitative measurements, crucial for assessing retinal layer thickness, edema, and the presence of subretinal fluids or lesions, thereby facilitating real-time retinal disease monitoring and diagnosis~\cite{retinareview2010,leandro2023oct}. Although OCT provides substantial advantages for retinal imaging, it faces limitations such as diminished clarity under certain conditions and speckle noise, which manifests as a grainy texture due to the spatial-frequency bandwidth limitations of interference signals. These limitations can lead to artifacts, often exacerbated by patient movement, potentially obscuring critical details necessary for accurate diagnosis and research. However, these speckle patterns are not just noise; they are thought to contain valuable information about the retinal tissue's microstructure~\cite{6556778}, which could be harnessed for detailed disease analysis and diagnosis.

In response to the need for a swift and non-invasive method of obtaining high-resolution, detailed confocal images, we turn to the burgeoning field of deep learning-based medical image-to-image translation (I2I)~\cite{chen2023deep}. 
I2I is employed to transfer multimodal medical images from one domain to another, aiming to synthesize less accessible but informative images from available images. The translation supports further analytical tasks, utilizing imaging modalities to generate images that are difficult to acquire due to invasiveness, cost, or technical limitations~\cite{wang2018conditional,liao2018adversarial, zhao2018towards, mahapatra2018efficient}. Thus, it enhances the utility of existing datasets and strengthens diagnostics in fields like ophthalmology~\cite{mahapatra2017image, abdelmotaal2022bridging}, where multimodal approaches have shown advantages over uni-modal ones in analysis and diagnosis of diabetic eye diseases (mainly DR), diabetic macular edema, and glaucoma~\cite{kalloniatis2024glaucoma,el2023eye, kang2021multimodal, el2024multimodality,hasan2024artificial}.
In OCT to Confocal translation, I2I aims to transfer information that is challenging to visualise in OCT images into the clear, visible confocal domain, preserving the structure of OCT while enriching them with high-resolution and cellular-level details. By learning the relationship between confocal microscopy cell distribution and OCT speckle patterns, we aim to synthesize "longitudinal confocal images", revealing information traditionally obscured in OCT. This advance aids early disease detection and streamlines treatment evaluation, offering detailed retinal images without the ethical concerns or high costs of conventional confocal methods.

Common medical image-to-image translation approaches have evolved significantly with the advent of Generative Adversarial Networks (GAN)~\cite{gan,shi2023translation}. For instance, the introduction of pix2pix~\cite{pix2pix}, a supervised method based on conditional GANs, leveraging paired image as a condition for generating the synthetic image. However, obtaining such paired images can be challenging or even infeasible in many medical scenarios. Consequently, unpaired image-to-image translation methods, like CycleGAN~\cite{cyclegan}, have emerged to fill this gap, addressing these limitations by facilitating the translation without the need for paired images. These methods have been successfully applied to modalities like MRI and CT scans~\cite{xia2022image, zhang2018translating, boulanger2021deep}, yet the challenge of translating between fundamentally different image domains, such as from 3D volumetric grayscale OCT to color confocal images at the cellular level remains relatively unexplored. Translations of this nature require not only volume preservation but also intricate cellular detail rendering in color, different from the grayscale to grayscale transitions typically seen in MRI-CT~\cite{GU2023107571} or T1-weighted and T2-weighted MRI conversions~\cite{welander2018generative}. This gap highlights the necessity for advanced translation frameworks capable of handling the significant complexity of OCT to confocal image translation, a domain where volumetric detail and cellular-level color information are both critical and yet to be thoroughly investigated.

In this paper, we propose a 3D modality transfer framework based on 3D CycleGAN to capture and transfer information inherent in OCT images to confocal microscopy. As registered ground truth is unavailable, the proposed framework is based on an unpaired training strategy. By extending the original CycleGAN approach, which processes 2D images slice-by-slice and often leads to spatial inconsistencies in 3D data, we incorporated 3D convolutions into our model. This adaptation effectively translates grayscale OCT volumes into rich, confocal-like coloured volumes, maintaining three-dimensional context for improved consistency and continuity across slices. We also unveil the OCT2Confocal dataset, a unique collection of unpaired OCT and confocal retinal images, poised to be the first of its kind for this application. This manuscript builds upon our initial investigation of this topic, with preliminary results presented in~\cite{oct2confocal}. In conclusion, the core contributions of our work are as follows:
\begin{enumerate}
    \item We introduce a 3D CycleGAN framework that first addresses the unsupervised image-to-image translation from OCT to confocal images. The methodology exploits the inherent information of in vivo OCT images and transfers it to ex vivo confocal microscopy domain without the need for paired data
    \item Our framework effectively captures and translates three-dimensional retinal textures and structures, maintaining volumetric consistency across slices. The result shows enhanced interpretability of OCT images by synthesizing confocal-like details, which may potentially aid improved diagnostic processes without the constraints of traditional methods
    \item The introduction of the OCT2Confocal dataset, a unique collection of OCT and confocal retinal images, which facilitates the development and benchmarking of cross-modal image translation research.
\end{enumerate}

The remainder of this paper is organized as follows: Section~\ref{sec:relatedwork} reviews relevant literature, contextualizing our contributions within the broader field of medical image translation. Section~\ref{sec:method} outlines our methodological framework, detailing the architecture of our 3D CycleGAN and the rationale behind its design. Section~\ref{sec:dataset} describes our novel OCT2Confocal dataset. Section~\ref{sec:experimental} presents the experimental setup, including the specifics of our data augmentation strategies, implementation details, and evaluation methods. Section~\ref{sec:result} presents the results and analysis with ablation studies, dissecting the impact of various architectural choices and hyperparameter tunings on the model's performance, quantitative metrics, and qualitative assessments from medical experts. Finally, Section~\ref{sec:conclusion} concludes with a summary of our findings and an outlook on future directions, including enhancements to our framework and its potential applications in clinical practice.

\section{Related Work}
\label{sec:relatedwork}

The importance of image-to-image translation is increasingly recognized, particularly for its applications ranging from art creation and computer-aided design to photo editing, digital restoration, and especially medical image synthesis~\cite{i2i}.

Deep generative models have become indispensable in this domain, with i) VAEs (Variational AutoEncoders)~\cite{vae} which encode data into a probabilistic latent space and reconstruct output from latent distribution samples, effectively compressing and decompressing data while capturing its statistical properties; ii) DMs (Diffusion Models)~\cite{ddpm,unsb} which are parameterized Markov chains, trained to gradually convert random noise into structured data across a series of steps, simulating a process that reverses diffusion or Brownian motion; and iii) GANs ~\cite{gan} which employ an adversarial process wherein a generator creates data in an attempt to deceive a discriminator that's trained to differentiate between synthetic and real data. VAEs often produce blurred images lacking in detail~\cite{zhao2017towards}, while DMs often fall short of the high standards set by GANs and are computationally slower~\cite{10223912}. GANs are particularly noted for their ability to generate high-resolution, varied, and style-specific images, making them especially useful in medical image synthesis~\cite{brock2018large, karras2020analyzing, medgan, wang2023bright, li2020generating}. In particular, those based on models such as StyleGAN~\cite{stylegan} and pix2pix~\cite{pix2pix} architectures, offer significant improvements in image resolution and variety, although with certain limitations. StyleGAN, an unconditional generative adversarial network, performs well within closely related domains but falls short when faced with the need for broader domain translation. On the other hand, pix2pix operates as a conditional GAN that necessitates paired images for the generation of synthetic images. While powerful, this requirement often poses significant challenges in medical scenarios where obtaining precisely pixelwise matched, paired datasets is difficult or sometimes impossible. 

Unpaired image-to-image translation methods, like CycleGAN~\cite{cyclegan}, emerged to address the need for paired datasets. CycleGAN, equipped with two generators and two discriminators (two mirrored GANs), enforces style fidelity by training each generator to produce images indistinguishable from the target domain by mapping the statistical distributions from one domain to another. It utilizes the cycle consistency loss~\cite{cycleloss} to ensure the original input image can be recovered after a round-trip translation (source to target and back to source domain) to preserve the core content. This architecture has shown effectiveness in biological image-to-image translation~\cite{bourou2023unpaired} and medical image-to-image translation tasks, such as MRI and CT scan translations~\cite{xia2022image, zhang2018translating, wang2021dicyc} and fluorescein angiography and retinography translations~\cite{hervella2019deep}, demonstrating its utility in scenarios where direct image correspondences are not available, and showing its capability of broader domain translation.

Notably, a significant gap remains in the translation of 3D medical images, where many existing methods simulate a 3D approach by processing images slice-by-slice rather than as complete volumes~\cite{welander2018generative, peng2023image}. While some work has been done in the 3D CycleGAN space, such as in translating between diagnostic CT and cone-beam CT (CBCT)~\cite{glloss}, these efforts have not ventured into the more complex task of translating between fundamentally different domains, such as from grayscale OCT images to full-color confocal microscopy. Such translations not only require the preservation of volumetric information but also a high-fidelity rendering of cellular details in color, distinguishing them from more common grayscale image-to-image translation.

In summary, translating OCT images into confocal is a novel problem in medical image-to-image translation. This process, which involves the translation from grayscale to full-color 3D data, has yet to be explored, particularly using a dedicated 3D network. This is the focus of our work.

\section{Proposed Methodology} 
\label{sec:method}
\subsection{Network Architecture}

\begin{figure}[!htb]
    \centering
    \includegraphics[width=0.9\textwidth]{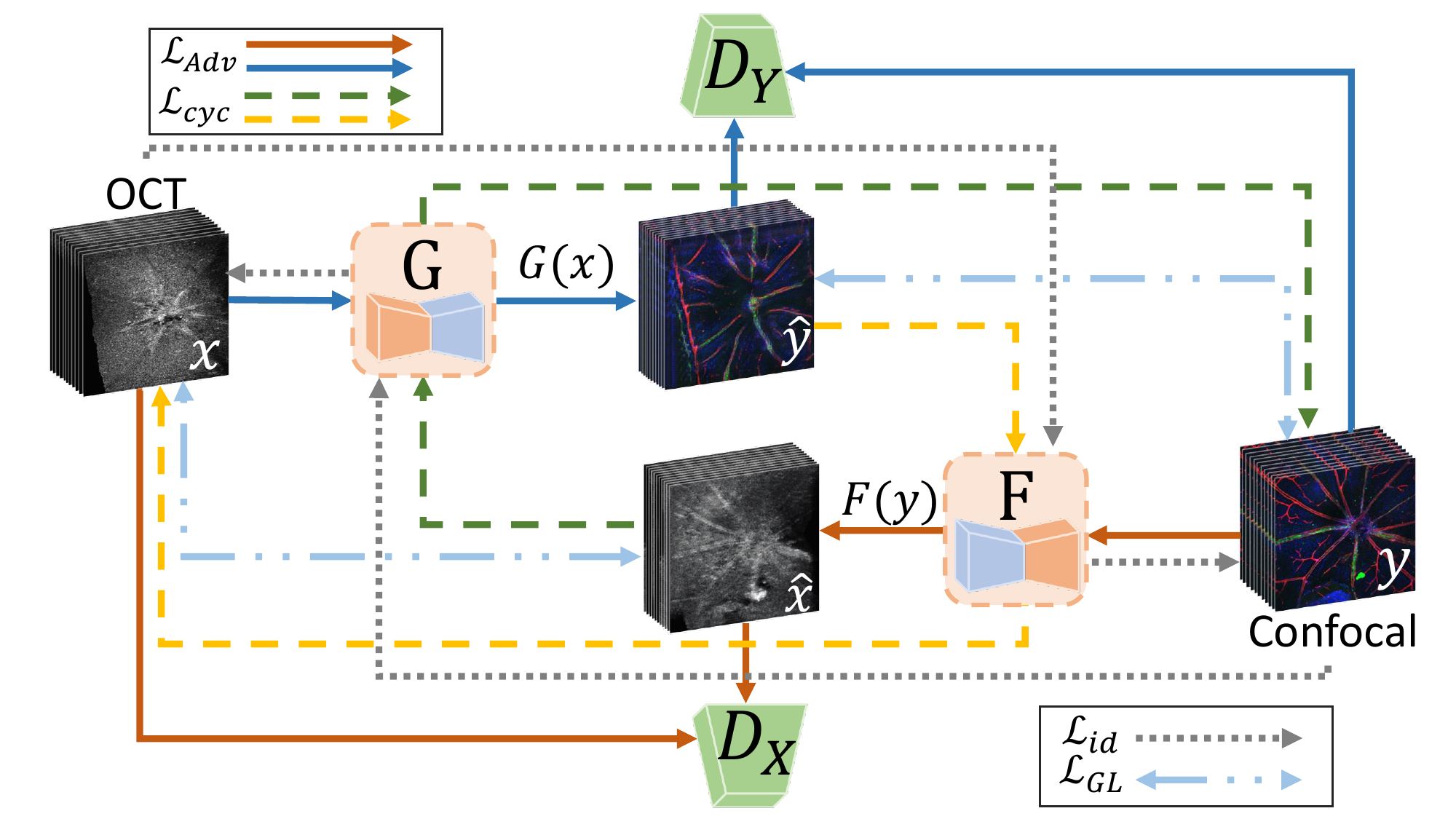}
    \caption{The proposed OCT-to-Confocal image translation method based on 3D CycleGAN}
    \label{fig:3dcyclegan}
    \end{figure}

The proposed 3D CycleGAN method, an extension of the 2D CycleGAN architecture~\cite{cyclegan}, employs 3D convolutions to simultaneously extract the spatial and depth information inherent in image stacks. Given an OCT domain $X$ and a Confocal domain $Y$, the aim of our model is to extract statistic information from both $X$ and $Y$ and then learn a mapping $G: X \to\ Y$ such that the output $\hat{y}=G(x)$, where $x \in X$ and $y \in Y$. An additional mapping $F$ transfers the estimated Confocal $\hat{y}$ back to the OCT domain $X$. The framework comprises two generators and two discriminators to map $X$ to $Y$ and vice versa. The input images are processed as 3D stacks, and all learnable kernels within the network are three-dimensional, as depicted in Figure~\ref{fig:3dcyclegan}.

\subsection{Generators and Discriminators of 3D CycleGAN}

\subsubsection{Generator}
The generator $G$ begins with a Convolution-InstanceNorm-ReLU layer, followed by three downsampling layers, and 9 residual blocks~\cite{resnet} that process the image features. It accepts an input of OCT cubes with dimensions \( H \times W \times D \times C_1 \), where \( H, W, \) and \( D \) represent height, width, and depth, respectively, and \( C_1 \) is the channel dimension, with \( C_1 = 1 \) indicating a single-channel grayscale format. Then, 3 fractional-strided convolution layers are used to increase the image size back to its original dimensions. Finally, the network concludes with a convolution layer that outputs the image in a 3-channel RGB format to construct confocal images, using reflection padding to avoid edge artifacts. Note that we have tested several settings, including U-Net architectures~\cite{unet}, WGAN-GP~\cite{wgan}, and the 9 residual blocks (ResNet 9) give the best results. The generator $F$ shares the identical architecture with generator $G$, but its final convolution layer outputs the image in a single channel to reconstruct OCT images. It processes input dimensions \( H \times W \times D \times C_2 \), where \( C_2 = 3 \) corresponds to the RGB color channels of the confocal microscopy images.


\subsubsection{Discriminator}
The discriminator networks in our framework are adaptations of the 2D PatchGAN~\cite{pix2pix} architecture. In our implementation, the 3D PatchGANs assess 70x70x9 voxel cubes from the 3D images to evaluate their authenticity. The key benefits of utilizing a voxel-level discriminator lie in its reduced parameter count relative to a full image stack discriminator.

\subsection{The Loss Function}
The objective consists of four terms: 1) adversarial losses~\cite{gan} for matching the distribution of generated images to the data distribution in the target domain, 2) a cycle consistency loss to prevent the learned mappings $G$ and $F$ from contradicting each other, 3) an identity loss to ensure that if an image from a given domain is transformed to the same domain, it remains unchanged, and 4) the gradient loss to enhance the textural and edge consistency in the translated images

\begin{enumerate}
\item [1)]\textbf{Adversarial Loss}
    
In our model, the adversarial loss is based on the Binary Cross-Entropy (BCE) loss, as used in traditional GANs~\cite{gan}. It adapts the style of the source domain to match the target by encouraging the generators to produce outputs that are indistinguishable from the target domain images and is defined as:
\begin{align*}
\mathcal{L}_{\text{Adv}}(G, D_{Y}) &= \mathbb{E}_{y\sim p_{\text{data}}(y)}\![-\log D_{Y}(y)] \! \\
&\!+ \mathbb{E}_{x\sim p_{\text{data}}(x)}\![-\log(1\!-\!D_{Y}(G(x)))], 
\tag{1}
\end{align*}
where \(G\) denotes the generator creating confocal images \(G(x)\) that aim to be indistinguishable from real confocal images in domain \(Y\), and \(D_Y\) represents the discriminator, distinguishing between actual confocal \(y\) and translated images \(G(x)\). The BCE loss measures the discrepancy between the discriminator's predictions and the ground truth labels using a logarithmic function, which can be more sensitive to changes when the discriminator is making a decision. We use an equivalent adversarial BCE loss for the mapping function $F: Y \to\ X$ and its discriminator $D_X$ as $\mathcal{L}_{Adv}(F, D_X)$ to maintain the adversarial relationship in both translation directions. The adversarial losses ensure the translated images conform to the stylistic characteristics of the target domain.

\item [2)]\textbf{Cycle consistency loss}

Cycle consistency loss~\cite{cycleloss}, defined in Equation~\eqref{eq:cycle_consistency}, ensures the network learns to accurately translate an image \(x\) from domain \(X\) to domain \(Y\) and back to \(X\) via mappings \(G\) and \(F\) (forward cycle) and vice versa for an image \(y\) (backward cycle), preserving the original image's integrity.
\begin{align*}
    \mathcal{L}_{\text{cyc}}(G,\! F)&=\mathbb{E}_{x\sim p_{\text{data}}(x)}[\Vert F(G(x))-x \Vert_{1}]\\ &+\mathbb{E}_{y\sim p_{\text{data}}(y)}[\Vert G(F(y))-y \Vert_{1}]. 
    \tag{2} 
    \label{eq:cycle_consistency}
\end{align*}

The $L_1$ loss between the original and translated backed image minimizes information loss, ensuring that the transformed image retains essential details and the core content of the input image.

\item [3)]\textbf{Identity Loss}
It was shown in~\cite{idloss} that adding identity losses can enhance the performance of the CycleGAN by preserving color consistency and other low-level information between the input and output, defined as follows:
\begin{align*}
\mathcal{L}_{\text{id}}(G,\! F) =& \ \mathbb{E}_{x\sim p_{\text{data}}(x)}[\Vert F(x)-x \Vert_{1}]\\
                                & +\mathbb{E}_{y\sim p{\text{data}}(y)}[\Vert G(y)-y \Vert_{1}].
\tag{3}
\end{align*}

The identity loss is calculated by taking the $L_1$ norm of the difference between a source domain image and its output after being passed through the generator designed for the opposite domain. For instance, if an OCT image is fed into a generator trained to translate confocal images to OCT images (opposite domain), the generator should ideally return the original OCT images unchanged. This process helps maintain consistent color and texture and indirectly stabilizes style fidelity.

\item [4)]\textbf{Gradient Loss}
The gradient loss promotes textural fidelity and edge sharpness by minimizing the $L_1$ norm difference between the gradients of real and synthesized images~\cite{glloss}, 
thereby preserving detail clarity and supporting both style rendering and information preservation through the enhancement of smooth transitions and the maintenance of edge details. The gradient loss is defined as:
\begin{align*}
    \mathcal{L}_{\text{GL}}(G,\ F) =& \ \mathbb{E}_{x\sim p_{\text{data}}(x)}[\Vert \nabla G(x)-\nabla y \Vert_{1}]\\
    & + \mathbb{E}_{y\sim p_{\text{data}}(y)}[\Vert \nabla F(y)-\nabla x \Vert_{1}],
\tag{4}
\end{align*}
where \(\nabla\) denotes the gradient operator. The term \(\nabla G(y) - \nabla y\) represents the difference between the gradients of the generated image \( G(y) \) and the real image \( y \).

\end{enumerate}

The total objective loss to minimize is the weighted summation of the four losses: the adversarial, the cyclic, the identity, and the gradient, given as follows:
\begin{align*}
\mathcal{L}_{\text{total}} = & \ \mathcal{L}_{\text{Adv}}(G,\ D_{Y}) \nonumber + \mathcal{L}_{\text{Adv}}(F, D_{X}) \nonumber \\
                             & + \lambda_1 \mathcal{L}_{\text{cyc}} + \lambda_2 \mathcal{L}_{\text{id}} +\lambda_3 \mathcal{L}_{\text{GL}}
\tag{3}
\end{align*}
where \(\lambda_1\), \(\lambda_2\) and \(\lambda_3\) are hyperparameters.

\section{OCT2Confocal Dataset} 
\label{sec:dataset}
\begin{figure}[!htb]
    \centering
    \includegraphics[width=\columnwidth]{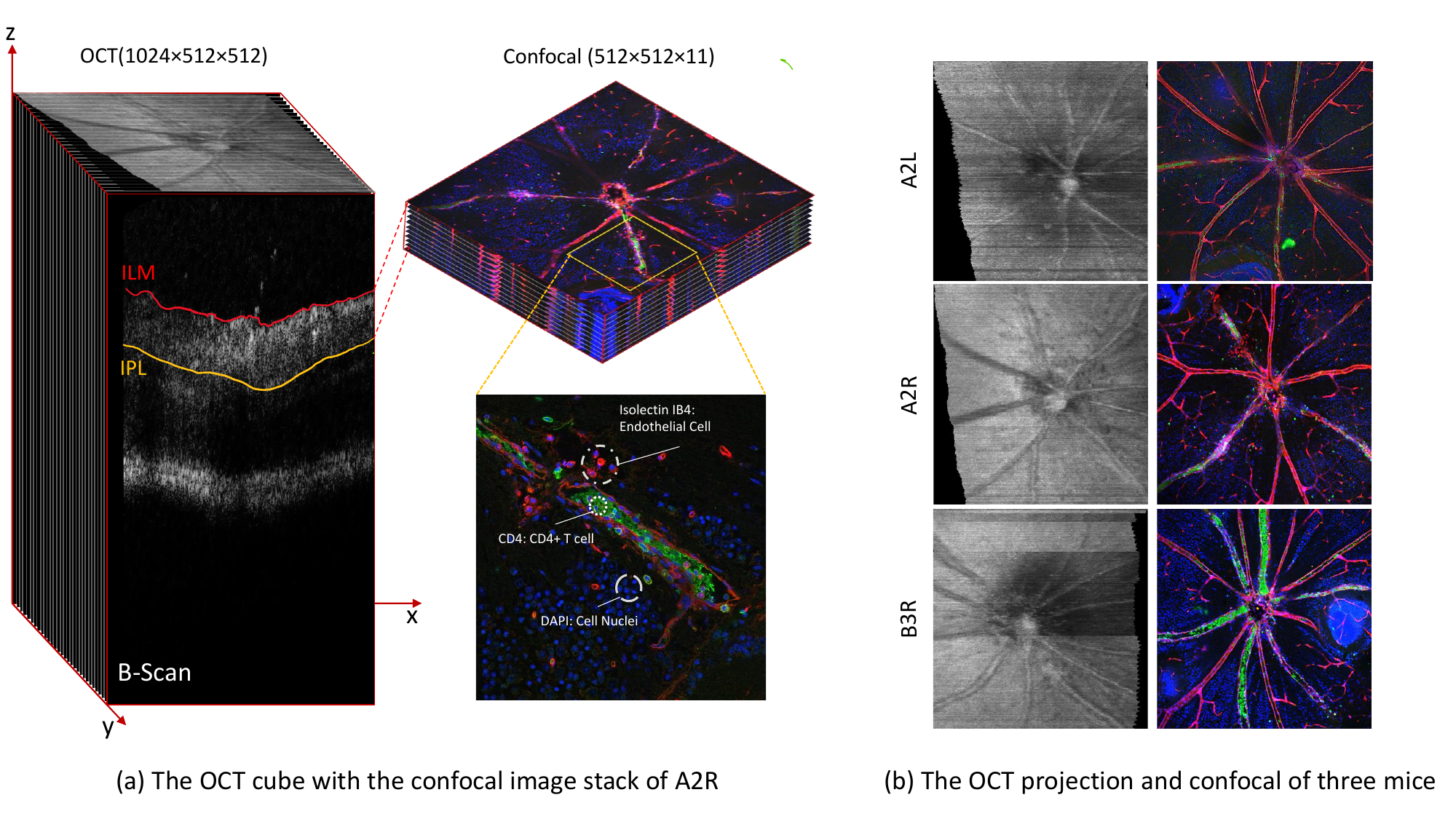}
    \caption{OCT2Confocal data. (a) The OCT cube with the confocal image stack of A2R, (b) The OCT projection and confocal of three mice}
    \label{fig:A2R}
    \end{figure}

We introduce the OCT2Confocal dataset, to the best of our knowledge, the first to include in-vivo grayscale OCT and corresponding ex-vivo colored confocal images from C57BL/6 mice, a model for human disease studies~\cite{boldison2015novel, caspi2010look}, with induced autoimmune uveitis. Our dataset specifically features three sets of retinal images, designated as A2L, A2R, and B3R. These identifiers represent the specific mice used in the study, with 'A2', 'B3' denoting the individual mice, and 'L' and 'R' indicating the left and right eyes, respectively. An example of the A2R data is shown in Figure~\ref{fig:A2R} (a). It is important to note that although the training data consists of 3D volumes, for the sake of clarity in visualization and ease of understanding, throughout the paper we predominantly display 2D representations of the OCT and confocal images (Figure~\ref{fig:A2R} (b)).

\begin{enumerate}
\item [a)] \textbf{The in-vivo Optical Coherence Tomography (OCT) images} were captured at various time points (days 10, 14, 17, and 24) using the Micron IV fundus camera equipped with an OCT scan head and a mouse objective lens provided by Phoenix Technologies, California. The resolutions of mice OCT images are 512$\times$512$\times$1024 (\( H \times W \times D \)) pixels, which is significantly smaller than human OCT images. Artefacts in OCT images, such as speckle noise and striped lines, can arise from motion artefacts, multiple scattering, attenuation artefacts, or beam-width artefacts. Volume scans, or serial B-scans (Figure~\ref{fig:A2R} (a)) defined at the x-z plane, were centered around the optic disc~\cite{retinareview2010}. In this study, for image-to-image translation from OCT to Confocal microscopy, the OCT volumetric data captured on day 24 is utilized to align with the day when confocal microscopy images are acquired. Specifically, the selected OCT volumes encompass the retinal layers between the Inner Limiting Membrane (ILM) and Inner Plexiform Layer (IPL) to align with the depth characteristics of the corresponding confocal microscopy images. The OCT B-scans are enhanced through linear intensity histogram adjustment and the Adaptive-Weighted Bilateral Filter (AWBF) denoising proposed by Anantrasirichai et al.~\cite{awbf}. The 2D OCT projection image, defined at the x-y plane (Figure~\ref{fig:A2R} (b)), is generated by summing up the OCT volume along the z-direction.





\item [b)] \textbf{The ex-vivo confocal image}. After the OCT imaging phase, the mice were euthanized on day 24, and their retinas were extracted and prepared for confocal imaging. The retinas were flat-mounted, and sequential imaging was performed using adaptive optics with a Leica SP5-AOBS confocal laser scanning microscope connected to a Leica DM I6000 inverted epifluorescence microscope. The retinas were stained with antibodies attached to four distinct fluorochromes, resulting in four color channels (Figure~\ref{fig:rawconfocal}):
    \begin{figure}[!htb]
    \centering
    \includegraphics[width=0.9\columnwidth]{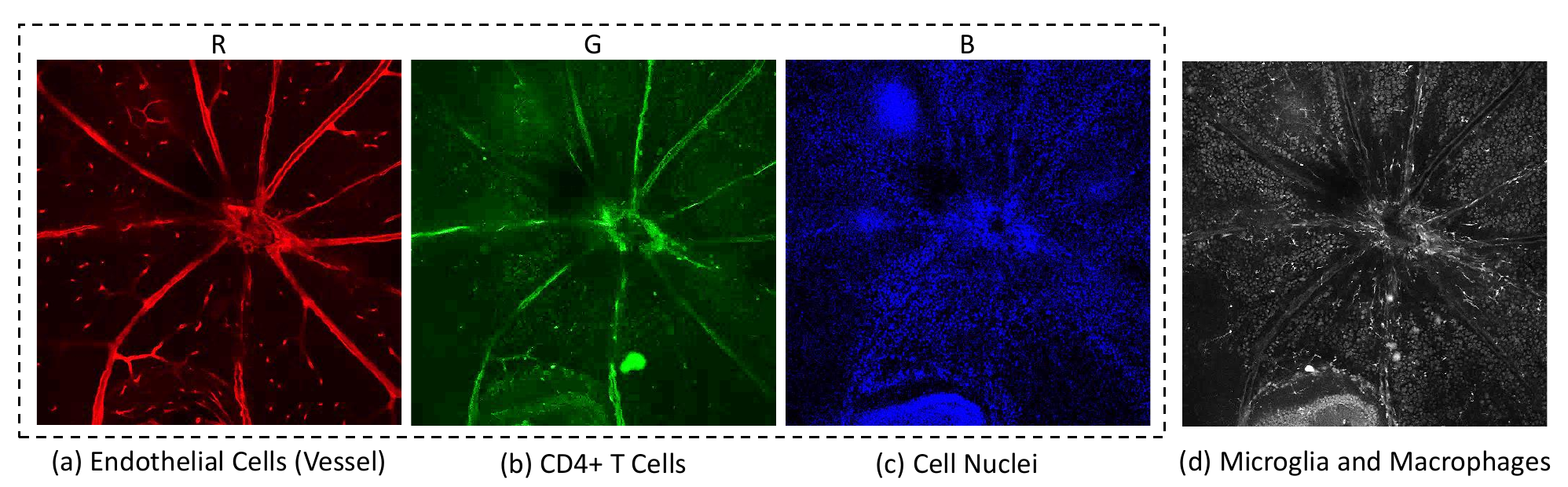}
    \caption{Example of one slice in an original four-color channel of retinal confocal image stack. The images show (from left to right):(a) Endothelial cells lining the blood vessels (red), (b) CD4+ T cells (green), (c) Cell nuclei stained with DAPI (blue), and (d) Microglia and macrophages (white)}
    \label{fig:rawconfocal}
    \end{figure}
    
\begin{itemize}

    \item Red (Isolectin IB4) channel (Figure~\ref{fig:rawconfocal} (a)), staining endothelial cells lining the blood vessels. This is important as changes in retinal blood vessels can indicate a variety of eye diseases such as diabetic retinopathy, glaucoma, and age-related macular degeneration.
    \item Green (CD4) channel (Figure~\ref{fig:rawconfocal} (b)), highlighting CD4+ T cells, which are critical in immune responses and can indicate an ongoing immune reaction in the retina.
    \item Blue (DAPI) channel (Figure~\ref{fig:rawconfocal} (c)), which stains cell nuclei, giving a clear picture of cell distribution.
    \item White (Iba1) channel (Figure~\ref{fig:rawconfocal} (d)), staining microglia and macrophages, providing insights into the state of the immune system in the retina.

\end{itemize}

This specific representation of cell types and structures via distinct color channels, referred to as the 'color code', is critical for the interpretability and utility of the confocal images in retinal studies. Specifically, the blue channel represents the overall cell distribution within the retina, the green channel highlights areas of immune response, and the red channel delineates the contour of the vessels. Thus, combining these three channels, we create an RGB image encompassing a broader range of retina-relevant information, forming the training set, and providing comprehensive colored cellular detail essential for the model training process. These RGB confocal images, with their corresponding Day 24 OCT images, were used for the training of the translation process. The confocal images include resolutions of A2L at 512×512×14 pixels, A2R at 512×512×11 pixels (shown in Figure~\ref{fig:A2R} (a)), and B3R at 512×512×14 pixels, all captured between the ILM and IPL layers. 

\end{enumerate}

Additionally, 23 OCT images without confocal matches from the retinal OCT dataset, also with induced autoimmune uveitis, introduced by Mellak et al.~\cite{mellak2023machine} were used to assess the model's translation performance as a test dataset. The OCT images are derived from either different mice or the same mice on different days, which also makes the dataset suitable for longitudinal registration tasks as performed in~\cite{3doctreg}. This OCT2Confocal dataset initiates the application of OCT-to-confocal image translation and holds the potential to deepen retinal analysis, thus improving diagnostic accuracy and monitoring efficacy in ophthalmology.


\section{Experimental Setup}
\label{sec:experimental}
\subsection{Dataset augmentation}

Our dataset expansion employed horizontal flipping, random zooming (0.9-1.1 scale), and random cropping, which aligns with common augmentation practices in retinal imaging~\cite{goceri2023medical}. Horizontal flipping is justified by the inherent bilateral symmetry of the ocular anatomy, allowing for clinically relevant image transformations. Random zoom introduces a controlled variability in feature size, reflecting physiologic patient diversity encountered in clinical practice. Random cropping introduces translational variance and acts as a regularization technique, mitigating the risk of the model overfitting to the borders of training images. These augmentation strategies were specifically chosen to avoid the introduction of non-physiological distortions that could potentially affect clinical diagnosis.


\subsection{Implementation Details}

The implementation was conducted in Python with the PyTorch library. Training and evaluation took place on the BlueCrystal Phase 4 supercomputer~\cite{bp} at the University of Bristol, featuring Nvidia V100 GPUs with 32 GB RAM, and a local workstation with RTX 3090 GPUs.

For our experiments, the OCT image cubes processed by generator \( G \) were sized \( 512 \times 512 \times 9 \times C_1 \), with \( C_1 = 1 \) indicating grayscale images. Similarly, the confocal images handled by generator \( F \) had dimensions \( 512 \times 512 \times 9 \times C_2 \), where \( C_2 = 3 \) represents the RGB color channels.

Optimization utilized the Adam optimizer~\cite{adam} with a batch size of 1, and a momentum term of 0.5. The initial learning rate was set at \(2 \times 10^{-5}\), with an input depth of 9 slices. Loss functions were configured with $\lambda_1= 8$, $\lambda_2= 0.1$, $\lambda_3 = 0.1$ . The 400-epoch training protocol maintained the initial learning rate for the first 200 epochs, then transitioned to a linear decay to zero over the next 200 epochs. Weights were initialized from a Gaussian distribution \( \mathcal{N}(0, 0.02) \), and model parameters were finalized at epoch 300 based on FID and KID performance.

\subsection{Evaluation Methods}
\subsubsection{Quantitative Evaluation}

The quantitative evaluation of image translation quality is conducted employing Distribution-Based (DB) objective metrics~\cite{medicaliqa} due to their ability to gauge image quality without necessitating a reference image. Specifically, the Fréchet Inception Distance (FID)~\cite{fid} and Kernel Inception Distance (KID) scores~\cite{kid} were utilized. 

These metrics are distribution-based, comparing the statistical distribution of generated images to that of real images in the target domain. Their widespread adoption in GAN evaluations underscores their effectiveness in reflecting perceptual image quality. FID focuses on matching the exact distribution of real images using the mean and covariance of features, which can be important for capturing the precise details in medical images and the correct anatomical structures with the appropriate textures and patterns. KID, on the other hand, emphasizes the diversity and general quality of the generated images without being overly sensitive to outliers ensuring that the generated images are diverse and cover the range of variations seen in real medical images. Lower FID and KID scores correlate with higher image fidelity. 
\begin{enumerate}
\item [a)] \textbf{Fréchet Inception Distance~\cite{fid}} is calculated as follows:
\begin{align*}
    FID(r, g) = \|\mu_r - \mu_g\|^2_2 + \mathrm{Tr}(\Sigma_r + \Sigma_g - 2(\Sigma_r\Sigma_g)^{\frac{1}{2}}),
    \tag{3}
\end{align*}
\noindent where $\mu_r$ and $\mu_g$ are the feature-wise mean of the real and generated images, respectively, derived from the feature vector set of the real image collection as obtained from the output of the Inception Net-V3~\cite{ICV3}. Correspondingly, $\Sigma_r$ and $\Sigma_g$ are the covariance matrices for the real and generated images from the same feature vector set. $\mathrm{Tr}$ denotes the trace of a matrix, and $\|\cdot\|_2$ denotes the $L_2$ norm. A lower FID value implies a closer match between the generated distribution and the real image distribution. Specifically in this study, the higher-dimensional feature vector sets characterized by 768-dimensional (FID768) and 2048-dimensional (FID2048) vectors are utilized as they capture higher-level perceptual and semantic information, which is more abstract and complex compared to the direct pixel comparison done by lower-dimensional feature spaces. These higher-dimensional features are likely to include important biomarkers and tissue characteristics critical for accurate image translation. 

\item [b)] \textbf{Kernel Inception Distance~\cite{kid}} is calculated using the Maximum Mean Discrepancy (MMD) with a polynomial kernel, as follows:
\begin{align*}
    KID(r, g) = & \frac{1}{m(m-1)}\sum_{i\neq j}k(x_i^r, x_j^r) \\
                & + \frac{1}{n(n-1)}\sum_{i\neq j}k(x_i^g, x_j^g) \\
                & - \frac{2}{mn}\sum_{i,j}k(x_i^r, x_j^g)
    \tag{3}
\end{align*}

where $m$ and $n$ are the numbers of real and generated images, respectively, $x_i^r$ and $x_j^g$ are the feature vectors of the real and generated images, respectively, and $k(x, y)$ is the polynomial kernel function.

\end{enumerate}

\subsubsection{Qualitative Evaluation}
The current objective metrics have been designed for natural images, limiting their performance when applied to medical imaging. Therefore, a subjective test leveraging a remote, crowd-based assessment was conducted for qualitative evaluation. This approach, contrasted with lab-based assessments, involved distributing the images to participants rather than hosting them in a controlled laboratory environment. The evaluation compared image-to-image translation results from five different methods: the UNSB diffusion model~\cite{unsb}, 2D CycleGAN, and three variations of the proposed 3D CycleGAN approach. This evaluation involved a panel of experts comprising 5 ophthalmologists and 5 individuals specializing in medical image processing. Participants were tasked with evaluating and ranking 5 images in their relative quality score for 13 sets of images. The 5 images in each set are resulted from the translation process of retinal OCT to confocal image translation. To mitigate sequence bias, the order of images within each set was randomized. Scores collected from the subjective testing were quantified and expressed as a Mean Opinion Score (MOS), which ranges from 1 to 100. Higher MOS values denote translations of greater authenticity and perceived quality. The evaluation was structured as follows:
\begin{itemize} 
    \item [1)] Initial Familiarization: The first 3 image sets included an original OCT image alongside its corresponding authentic confocal image and 5 translated confocal images from different methods and models, referred to as the With Reference (W Ref) group. These were provided to acquaint the participants with the defining features of confocal-style imagery.

    \item [2)] Blind Evaluation: The subsequent ten sets presented only the original OCT and 5 translated confocal images, omitting any genuine confocal references to ensure an unbiased assessment of the translation quality, referred to as the Without Reference (W/O Ref) group.

\end{itemize}
The participants were instructed to rank the images based on the following criteria:
\begin{itemize}
\item Authenticity: The degree to which the translated image replicates the appearance of a real confocal image.

\item  Color Code Preservation: Participants were advised to focus on the accuracy of color representation indicative of high-fidelity translation which are: i) The presence of green and red color to represent different cell types, with blue indicating cell nuclei, ii) The delineation of vessels by red, with green typically enclosed within these regions, iii) The alternation of green in vessels, where a green vessel is usually adjacent to non-green vessels, iv) The co-occurrence of red and green regions with blue elements.

\item Overall Aesthetic: The visual appeal of the image as a whole was also considered.

\item Artifact Exclusion: Any artifacts that do not impact the justification of overall image content should be overlooked.
\end{itemize}

Additionally, to substantiate the reliability of selected metrics(FID768, FID2048, and KID) for evaluating OCT to confocal image translations against MOS, Spearman's Rank-Order Correlation Coefficient (SROCC) and Linear Correlation Coefficient (LCC) were applied to both selected DB metrics and a range of No-Reference (NR) metrics, including FID64, FID192, FID768, FID2048, KID, NIQE~\cite{niqe}, NIQE\_M (a modified NIQE version trained specifically with parameters from the original confocal image dataset), and BRISQUE~\cite{brisque}. Both SROCC and LCC range from $-1$ to $+1$, where $+1$ indicates a perfect positive correlation, 0 denotes no correlation, and $-1$ signifies a perfect negative correlation. These analyses correlate the metrics with the MOS to assess the consistency and predictive accuracy of FID and KID in reflecting subjective image quality assessments.

\begin{table}[htb]
\centering
\caption{Correlation of selected DB and NR image quality metrics with MOS}
\begin{minipage}{\textwidth} 
\centering
\label{tab:srocc}
\begin{tabular}{@{}lcccccccc@{}}
\toprule
 & FID64 & FID192 & FID768 & FID2048 & KID & NIQE & NIQE\_M & BRISQUE \\ \midrule
SROCC & -0.3768 & -0.4235 & -0.7666 & -0.7823 & -0.8271 & 0.6346 & -0.5416 & 0.032 \\
LCC & -0.699 & -0.6894 & -0.7872 & -0.7813 & -0.8099 & 0.5955 & -0.5804 & -0.0446 \\ \bottomrule
\end{tabular}%
\footnotetext{Note: FID and KID metrics assess image quality, with lower values indicating better quality. NIQE and BRISQUE are no-reference image quality evaluators; lower NIQE scores suggest better perceptual quality, whereas BRISQUE evaluates image naturalness. SROCC and LCC measure the correlation between objective metrics and subjective MOS ratings. SROCC and LCC values closer to -1 or 1 indicate a strong correlation, with positive values suggesting a direct relationship and negative values an inverse relationship.}
\vspace{.15in}
\end{minipage}
\end{table}

From Table~\ref{tab:srocc}, the negative correlation of FID and KID metrics with MOS, as indicated by their SROCC values, aligns with the expectation for lower-the-better metrics. Notably, KID demonstrates the strongest negative correlation (-0.8271), closely followed by FID2048 (-0.7823) and FID768 (-0.7666), suggesting their effectiveness in reflecting perceived image quality. Conversely, NIQE's positive correlation (0.6346) contradicts this principle, questioning its suitability, while the modified NIQE\_M shows some improvement with a negative correlation (-0.5416). BRISQUE's low positive correlation (0.032) indicates a nearly negligible relationship with MOS. LCC results reinforce these findings, particularly highlighting KID's superior correlation (-0.8099). These analyses collectively suggest that KID, FID768, and FID2048 are relatively the most reliable metrics for evaluating the quality of translated Confocal images in this context, while the results for NIQE and BRISQUE imply limited applicability.

\section{Results and Analysis} 
\label{sec:result}
In this section, we analyze our proposed model through ablation studies and compare it with baseline methods both quantitatively and qualitatively. For clearer visualization, results are displayed as fundus-like 2D projections from the translated 3D volume. 

The ablation study investigates the impact of different generator architectures, hyperparameters of loss functions, and the number of input slices on our model's performance. This study is essential for understanding how each component contributes to the efficacy of the 3D CycleGAN framework in translating OCT to confocal images.

We compare our model against the UNSB diffusion model~\cite{unsb} and the conventional 2D CycleGAN~\cite{cyclegan}, underscoring the effectiveness of the GAN architecture and 3D network. 
As the UNSB and 2D CycleGAN are 2D models, 3D OCT and confocal images are processed into 2D slices along the z-direction, which are then individually translated and subsequently reassembled back into a 3D volume. This process allows us to directly compare the efficacy of 2D translation techniques on 3D data reconstruction. We evaluated against 3D CycleGAN variants: one with two downsampling layers (3D CycleGAN-2) without Gradient Loss, and another with the same layers but including Gradient Loss (3D CycleGAN-GL). Our final model, 3D CycleGAN-3 with three downsampling layers and gradient loss is also included in these comparisons. Each model is retrained on the same datasets and configurations for consistency. 




\subsection{Ablation Study}
\subsubsection{Generator Architecture}
\begin{figure}[htb]
\includegraphics[width=\columnwidth]{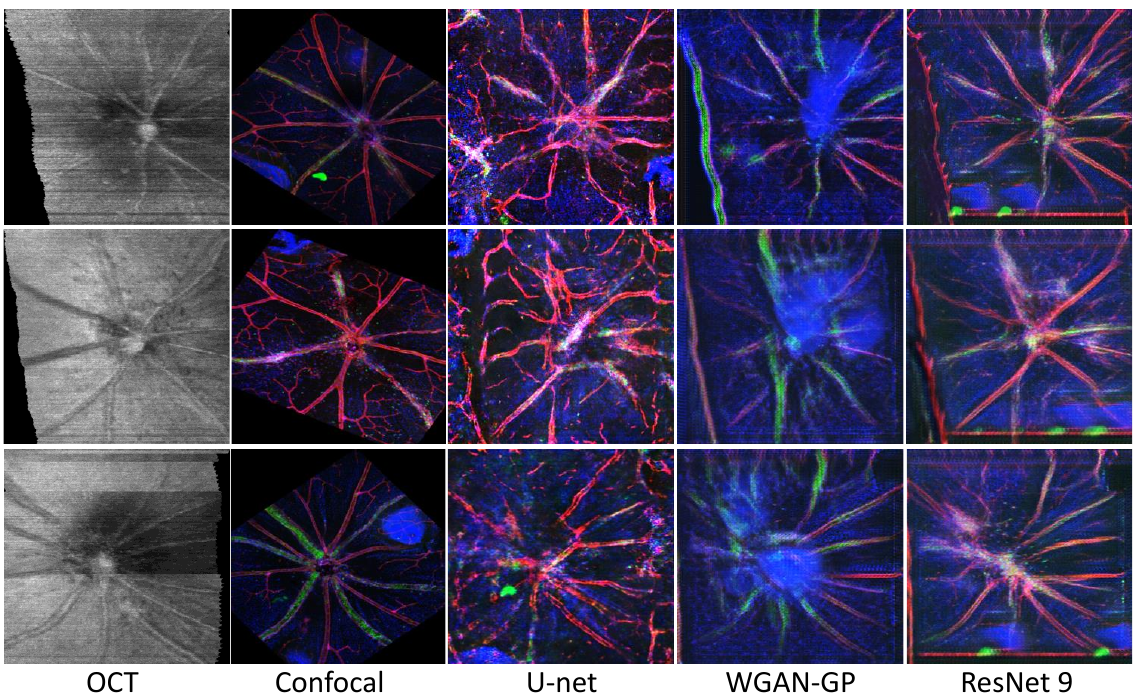}
\caption{Visual comparison of translated images using different generator architectures. This figure displays the translated confocal images using U-Net, WGAN-GP, and ResNet 9 architectures}
\label{fig:ablation}
\end{figure}

In our experiments, the structure of the generator is found to have the most significant impact on the generated results, overshadowing other factors such as the hyperparameters of gradient loss and identity loss. From Table~\ref{tab:ablation}, the ResNet 9 configuration emerges as the most effective structure, outperforming both U-Net and WGAN-GP models. The ResNet 9's lower FID scores suggest a superior ability to produce high-quality images that more closely resemble the target confocal domain. While WGAN-GP attains the lowest KID score, visual assessment in Figure~\ref{fig:ablation} shows that it still produces significant artefacts, underscoring the limitation of WGAN-GP and the limitation of FID and KID metrics assessing image quality in medical imaging contexts. On the other hand, the U-Net architecture, although commonly used for medical image segmentation, falls short in this generative task, particularly in preserving the definition and complex anatomical structures such as blood vessels and positions of the optic disc (where blood vessels converge), as shown in the second column of Figure~\ref{fig:ablation}. Meanwhile, the ResNet 9 maintains spatial consistency and detail fidelity, ensuring that synthesized images better preserve critical anatomical features, which is paramount in medical diagnostics.

\begin{table}[htb]
\caption{Comparative results of different generators architecture in 3D CycleGAN-3. The table presents FID768, FID2048, and KID scores for U-Net, WGAN-GP, and ResNet 9 generators. Lower scores indicate better performance, with the best result colored in {\color[HTML]{FF0000} red}.}
\begin{minipage}{\textwidth} 
\centering
\begin{tabular}{@{}lccc@{}}
\toprule
Generator & FID768 ↓ & FID2048 ↓ & KID ↓ \\ \midrule
U-Net & 1.135 & 178.445 & 0.182 \\
WGAN-GP & 1.202 & 173.142 & {\color[HTML]{FE0000}0.129} \\
ResNet 9 & {\color[HTML]{FE0000} 0.785} & {\color[HTML]{FE0000} 151.302} &  0.143 \\ \bottomrule
\end{tabular}%
\footnotetext{Note: FID768 and FID2048 refer to the Fréchet Inception Distance computed with 768 and 2048 features, respectively. KID refers to the Kernel Inception Distance. Both FID and KID indicate better performance with lower scores.}
\vspace{.15in}
\label{tab:ablation}
\end{minipage}
\end{table}

\subsubsection{Impact of Gradient and Identity Loss Hyperparameters}

\begin{figure}[htb]
\includegraphics[width=\columnwidth]{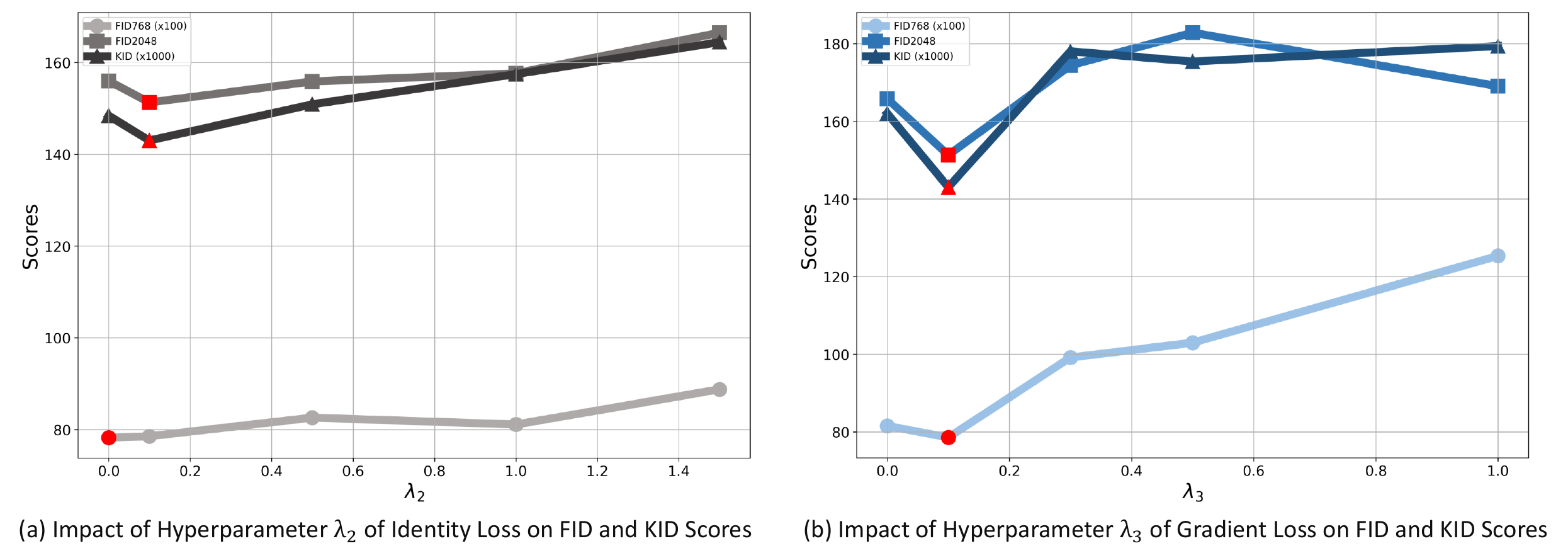}
\caption{Impact of Gradient and Identity Loss Hyperparameters \(\lambda_2\) and \(\lambda_3\) on FID and KID. The lowest (optimal) score is highlighted in {\color[HTML]{FF0000} red}}
\label{fig:hyperpaprameter}
\end{figure}

In our evaluation of the impact of identity loss (\(\lambda_2\)) and gradient loss (\(\lambda_3\)), we explore a range of values: \(\lambda_2\) at 0, 0.1, 0.5, and 1.5, and \(\lambda_3\) at 0, 0.1, 0.3, and 1.0. The line graphs in Figure~\ref{fig:hyperpaprameter} illustrate how these values affect the FID and KID scores, with the optimal balance achieved at 0.1 for both parameters, where the fidelity, the textural and edge details from the original OCT domain the target confocal domain is balanced.

\begin{figure}[htb]
\includegraphics[width=\columnwidth]{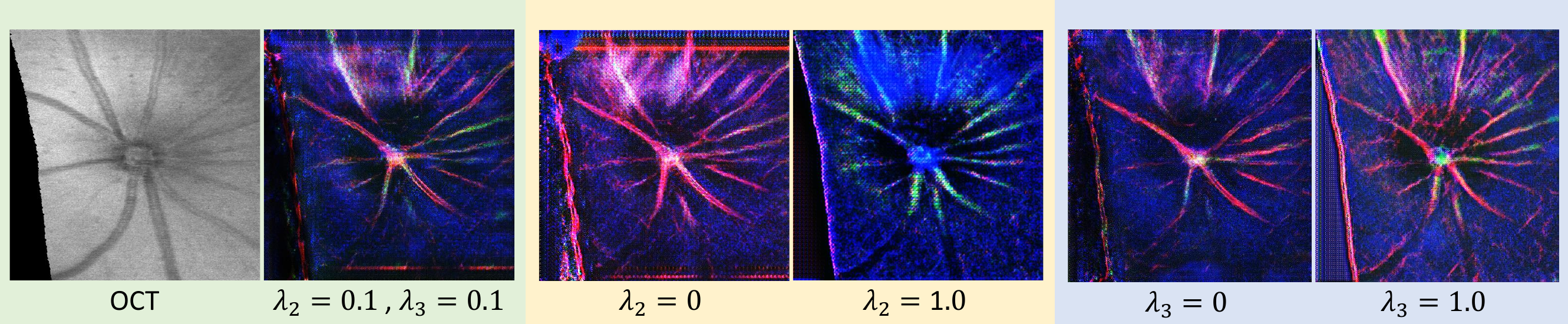}
\caption{Visual comparison of translated confocal images with different \(\lambda_2\) and \(\lambda_3\) values against the optimized setting}
\label{fig:lambda23}
\end{figure}

We observe that the absence of identity loss (\(\lambda_2 = 0\)), as visualized in Figure~\ref{fig:lambda23}, sometimes results in color misrepresentation in the translated images, such as pervasive blue or absent green hues, underscoring its role in maintaining accurate color distribution. In contrast, overemphasizing identity loss (\(\lambda_2 = 1.0\)) could lead to the over-representation of specific colors, raising the likelihood of artifacts.

 Similarly, without the gradient loss (\(\lambda_3 = 0\)), as shown in Figure~\ref{fig:lambda23}, some images exhibit a loss of detail, particularly blurring the delineation of cellular and vascular boundaries. Conversely, an excessive gradient loss (\(\lambda_3 = 1.0\)) overemphasizes minor vessels in the background and over-sharpens structures, occasionally distorting primary vessels.

In conclusion, the identity loss and the gradient loss are two losses with small but significant weights that help the model to focus on important essential features without causing an overemphasis that could detract from the overall image quality for OCT-to-confocal image translation.

\subsubsection{Impact of the Input Number of Slices of OCT and Confocal Images}

\begin{figure}[htb]
\includegraphics[width=\columnwidth]{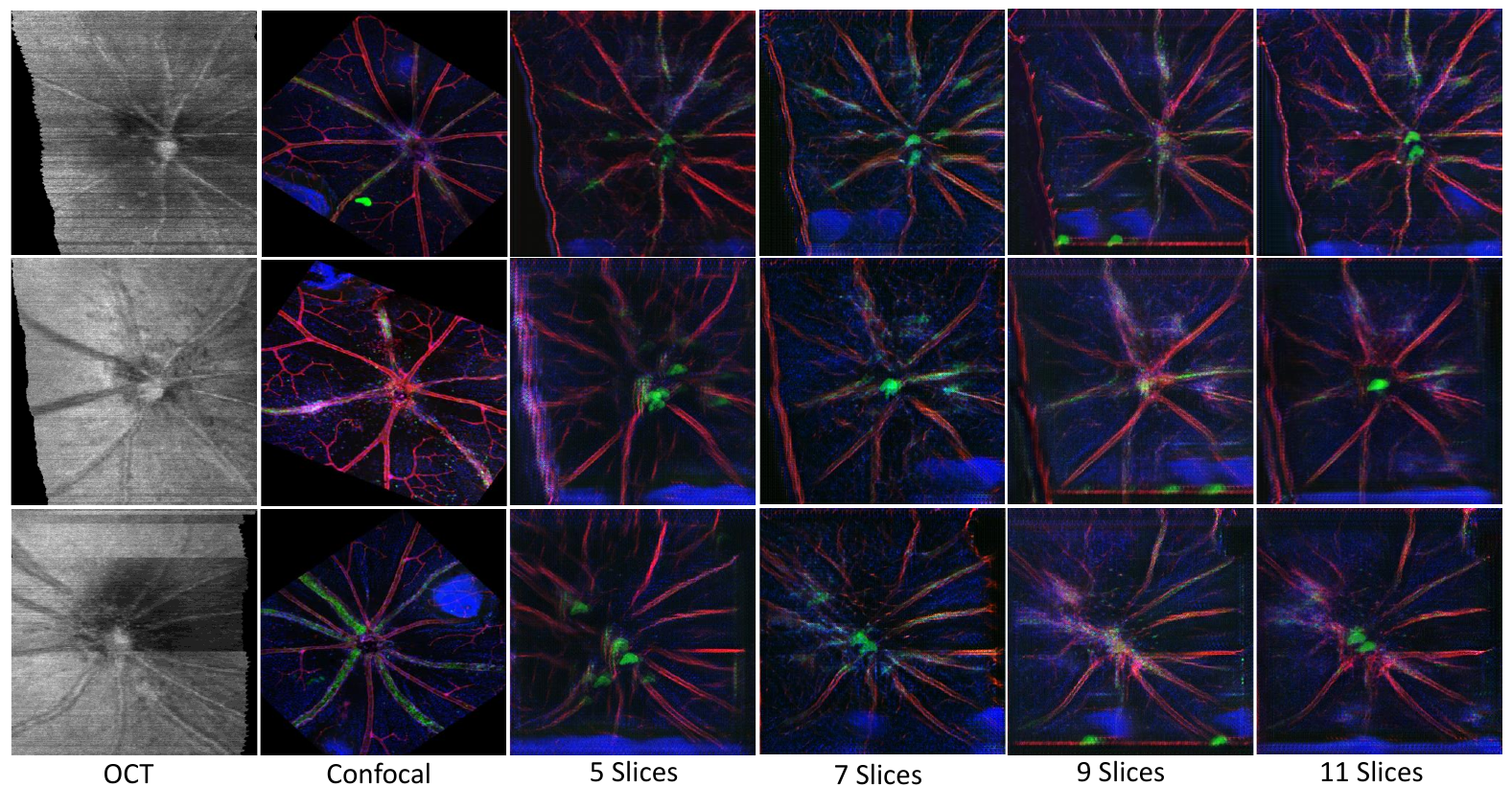}
\caption{Visual comparison of translated images with varying input slice depths (5, 7, 9, 11 slices). This figure demonstrates the impact of different slice depths on the quality of image translation by the 3D CycleGAN-3 model}
\label{fig:slicenumber}
\end{figure}

In our assessment of the 3D CycleGAN model's performance with different numbers of input slices (depth) for OCT and confocal images, we experimented with 5, 7, 9, and 11 slices. Due to the limited correlation between FID and KID metrics with visual quality across different slice counts, we primarily relied on visual assessments, as detailed in Figure~\ref{fig:slicenumber}.

Our findings indicate that at a depth of 5 slices, the model frequently exhibited repetitive artifacts and blocky textures, struggling to accurately map the color distribution from confocal to OCT images, which resulted in spatial inconsistencies and shadowing effects on blood vessels. Increasing the slice count to 7 improved color code preservation, yet issues with background shadowing remained, likely due to persisting spatial discrepancies. The optimal outcome is achieved with 9 slices, which effectively represented cell color distribution and maintained edge details, with minimal artifacts confined to less critical areas such as image borders. Although 11 slices theoretically should provide further improvements, it did not significantly outperform the 9-slice input and sometimes introduced central image artifacts. Considering computational efficiency and image quality, an input depth of 9 slices is selected as the standard for our model.

\subsection{Quantitative Evaluation}
In Table~\ref{tab:result}, we present both the DB image quality assessment results and subjective scores of the 13 selected OCT images set used in the subjective test. Across all DB metrics, the 3D CycleGAN-3 model outperformed other methods, achieving the lowest FID and KID scores in all scenarios (With Reference, Without Reference, and Total dataset). These results suggest that this model is most effective in aligning the statistical distribution of generated images with those of real images, indicating higher image fidelity and better perceptual quality. The 3D CycleGAN-2 model follows as the second best, performing notably well in the with-reference scenario. This suggests that the additional complexity of a third downsampling layer in 3D CycleGAN-3 does confer an advantage. An inference is that an extra downsampling layer in a 3D convolutional network improves feature extraction by broadening the receptive field, enabling the model to better discern and synthesize the key structural elements within volumetric medical images. Overall, the 3D CycleGAN models outperform the UNSB diffusion model and the 2D CycleGAN, demonstrating the inadequacy of the diffusion model based UNSB for translating OCT to confocal images and illustrating the limitations of 2D models when dealing with volumetric data.
 
\begin{table}[htb]
\centering
\caption{The performance of models evaluated by DB metrics FID scores and KID scores, alongside the subjective MOS rating. The results are referred to categories with reference ('W Ref'), without reference ('W/O Ref'), and total (`Total') image sets. For each column, the best result is colored in {\color[HTML]{FF0000} red} and the second best is colored in {\color[HTML]{0922DB} blue}.}
\begin{minipage}{\textwidth} 
\centering
\resizebox{\textwidth}{!}{%
\label{tab:result}
\begin{tabular}{@{}lcccccccccccc@{}}
\toprule
 & \multicolumn{4}{c}{W Ref} & \multicolumn{4}{c}{W/O Ref} & \multicolumn{4}{c}{Total} \\ \cmidrule(lr){2-5} \cmidrule(lr){6-9} \cmidrule(lr){10-13}  
\multirow{-2}{*}{Model} & FID768 ↓ & FID2048 ↓ & KID ↓ & MOS ↑ & FID768 ↓ & FID2048 ↓ & KID ↓ & MOS ↑ & FID768 ↓ & FID2048 ↓ & KID ↓ & MOS ↑ \\ \midrule
UNSB & 1.659 & 313.189 & 0.597 & 29.300 & 1.611 & 301.666 & 0.655 & 25.360 & 1.622 & 304.325 & 0.641 & 26.269 \\
2D CycleGAN & 1.547 & 225.302 & 0.300 & 36.667 & 1.420 & 231.048 & 0.326 & 41.630 & 1.449 & 229.722 & 0.320 & 40.485 \\
3D CycleGAN-GL & 1.281 & 202.795 & 0.267 & 50.400 & 1.170 & 169.556 & 0.215 & 49.550 & 1.195 & 177.227 & 0.227 & 49.746 \\
3D CycleGAN-2 & {\color[HTML]{0922DB} 0.852} & {\color[HTML]{FF0000} 149.486} & {\color[HTML]{FF0000} 0.144} & {\color[HTML]{0922DB} 53.967} & {\color[HTML]{0922DB} 0.890} & {\color[HTML]{0922DB} 166.473} & {\color[HTML]{0922DB} 0.160} & {\color[HTML]{0922DB} 52.860} & {\color[HTML]{0922DB} 0.881} & {\color[HTML]{0922DB} 162.553} & {\color[HTML]{0922DB} 0.156} & {\color[HTML]{0922DB} 53.115} \\
3D CycleGAN-3 & {\color[HTML]{FF0000} 0.766} & {\color[HTML]{0922DB} 154.756} & {\color[HTML]{0922DB} 0.153} & {\color[HTML]{FF0000} 56.867} & {\color[HTML]{FF0000} 0.780} & {\color[HTML]{FF0000} 155.188} & {\color[HTML]{FF0000} 0.156} & {\color[HTML]{FF0000} 56.350} & {\color[HTML]{FF0000} 0.777} & {\color[HTML]{FF0000} 155.089} & {\color[HTML]{FF0000} 0.155} & {\color[HTML]{FF0000} 56.469}\\
\bottomrule
\end{tabular}%
}
\footnotetext{Note: FID768 and FID2048 refer to the Fréchet Inception Distance computed with 768 and 2048 features, respectively. KID refers to the Kernel Inception Distance. Both FID and KID indicate better performance with lower scores. Mean Opinion Score (MOS) rates the subjective quality of images with higher scores reflecting better quality.}
\vspace{.15in}
\end{minipage}
\end{table}

\subsection{Qualitative Evaluation}

The 3D CycleGAN-3 model, as shown in Table~\ref{tab:result}, scored the highest in the MOS across all three scenarios as determined by the expert panel's rankings. This reflects the model's superior performance in terms of authenticity, detail preservation, and overall aesthetic quality. Notably, it also minimizes the presence of non-impactful artifacts, which is critical for the utility of translated images in clinical settings.

\begin{figure*}[htb]
\includegraphics[width=\textwidth]{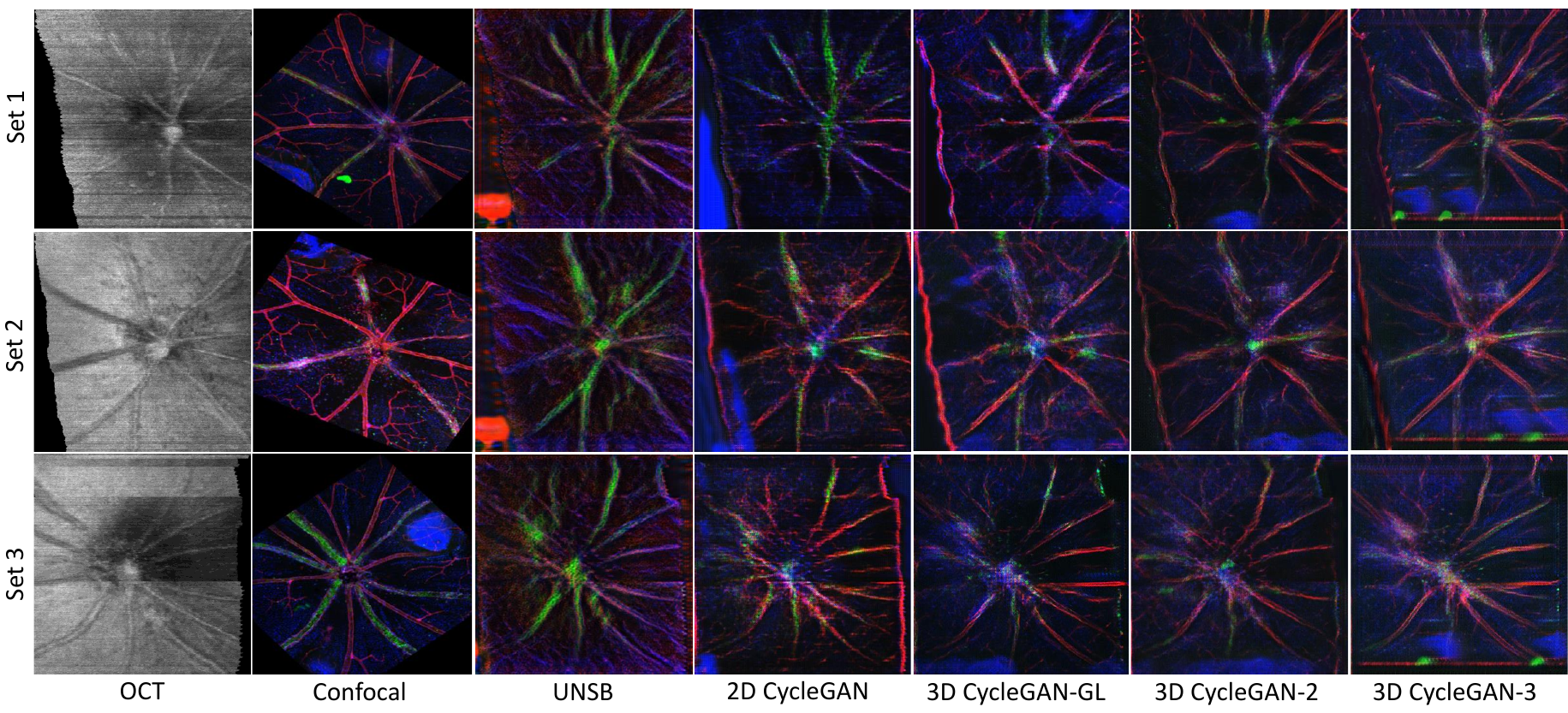}
\caption{Visual comparative translation results with reference}
\label{fig:wref}
\end{figure*}

\begin{figure*}[htb]
\includegraphics[width=\textwidth]{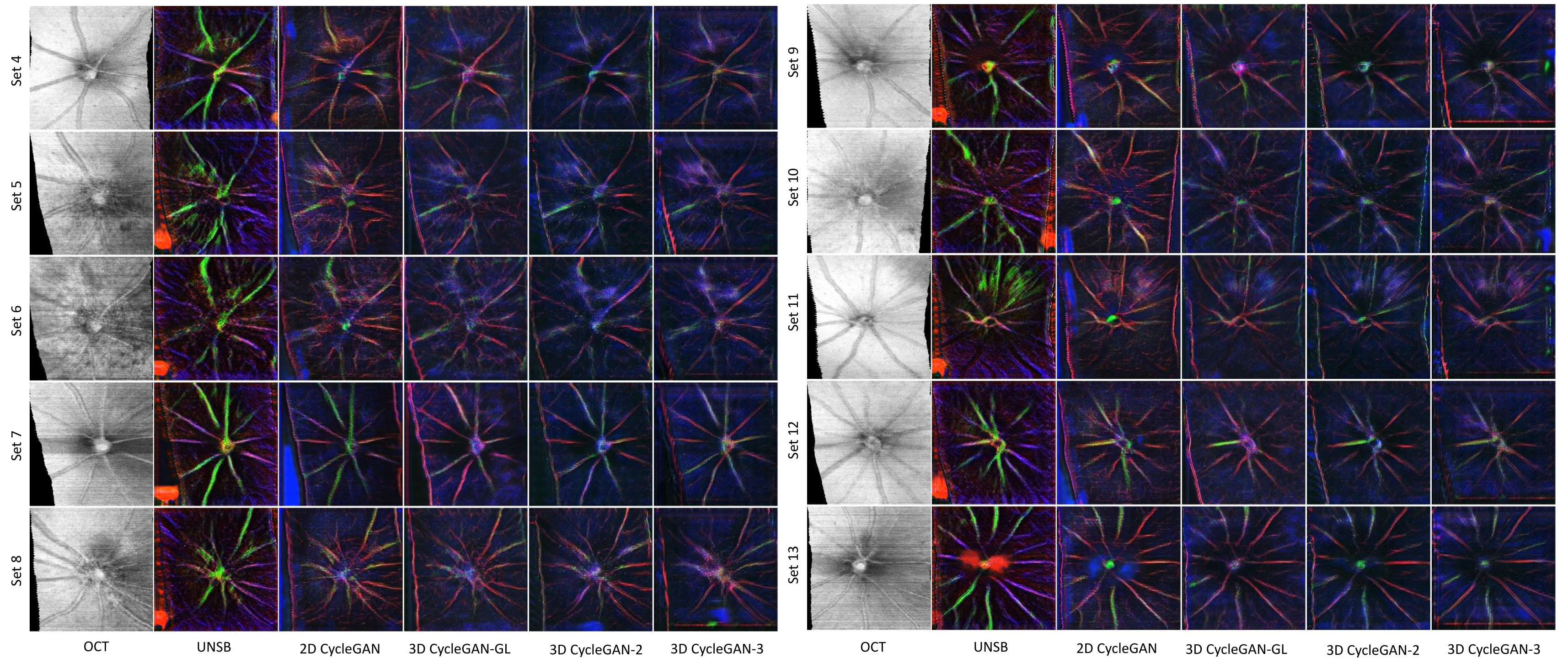}
\caption{Visual comparative translation results without reference}
\label{fig:woref}
\end{figure*}

\paragraph{Subjective Test} Analysis based on MOS and visual observations from Figure~\ref{fig:wref} and Figure~\ref{fig:woref} indicates that all 3D CycleGAN models effectively preserve blood vessel clarity, shape, and color code. The 3D CycleGAN-3 model, which received the highest MOS ratings in all scenarios, is reported to reflect the capacity for retaining more background detail and overall authenticity. Particularly in translating lower quality in-vivo OCT images (e.g., Set 6 in Figure~\ref{fig:woref}), the 3D CycleGAN-3 model demonstrates superior performance, highlighting its effectiveness in capturing the complex relationships between OCT and confocal domains. 

In contrast, the 2D models (2D CycleGAN and UNSB) sometimes introduce random colors, disregard edges, and inaccurately replicate retinal vessel color patterns. The UNSB, as a diffusion model, theoretically can generate diverse outputs. However, as indicated by its lower MOS scores and observed in Figure~\ref{fig:wref} and Figure~\ref{fig:woref}, it struggles significantly with preserving accurate color codes and structural details, leading to reduced visual quality and clinical usability in OCT to confocal translation. Conversely, CycleGAN-based models employ adversarial training to directly learn the transformation of input images into the target domain. This method is better at maintaining continuity in image quality and structure, providing visual advantages over the UNSB model. However, when compared to 3D models, these advantages diminish.

Specifically, when compared to the 3D CycleGAN-3, reconstructions from the 2D CycleGAN exhibit significant issues:  assembling 2D-processed images back into 3D often results in discontinuities in blood vessels and features across slices (z-direction). It manifests as repeated artifacts and features at various locations across different slices (xy-plane) and duplicated structures in 2D projections. As observed in Figure~\ref{fig:wref} (Set 2) and Figure~\ref{fig:woref} (Sets 6 and 11), the 2D CycleGAN results in more visible hallucinations than 3D CycleGAN-3 in the 2D projection images, where green artifacts occur at the optic disc (the convergence point for blood vessels). Moreover, numerous fine, hallucinated vascular structures appear in the background beyond the main vascular structures, which are absent in both the original OCT and confocal images, underscoring the limitations of 2D CycleGAN in handling the complexity of 3D data structures and maintaining spatial consistency.

\begin{figure}[htb]
\centering
\includegraphics[width=0.95\textwidth]{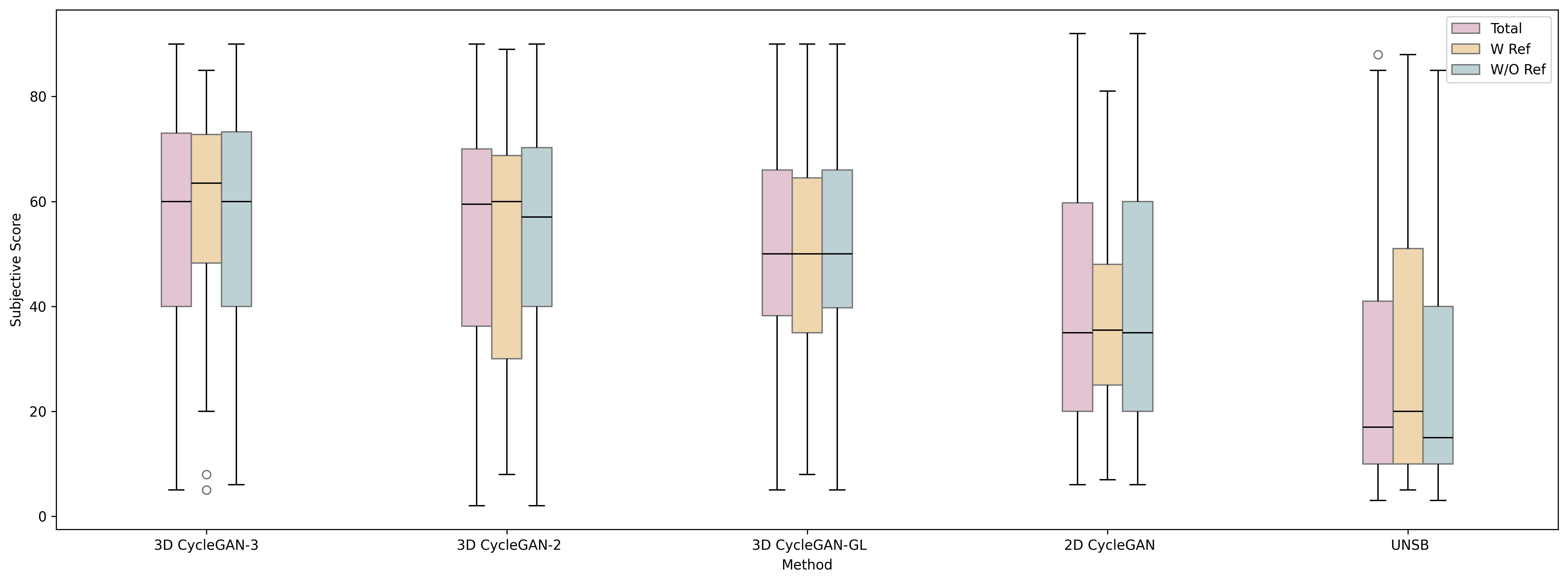}
\caption{Boxplot of subjective evaluation scores for comparison across scenarios with reference (`W Ref'), without reference (`W/O Ref'), and the combined total (`Total'). The circles indicate outliers in the data}
\label{fig:mos}
\end{figure}

Figure~\ref{fig:mos} presents a boxplot of MOS for the five evaluated methods, where 3D CycleGAN models outperform 2D models in translating OCT to confocal images. Specifically, the 3D CycleGAN-3 exhibits a more concentrated distribution of scores in MOS, indicating a consensus among experts on the quality of the generated confocal images by this model, underlining its proficiency in producing consistent and reliable translations. The statistical analysis conducted via Kruskal-Wallis tests across each scenario confirms significant differences among the methods (\( p < 0.001 \)). Subsequent pairwise Mann-Whitney U tests with Bonferroni adjustments clearly demonstrate that the 3D CycleGAN-3 model significantly outperforms both the 2D CycleGAN and UNSB models in all scenarios evaluated. For more detailed qualitative and quantitative results, please refer to Appendix \ref{appendixA}, where Table~\ref{tab:eachset} presents FID, KID, and MOS scores for each set evaluated in the subjective test.

\paragraph{Ophthalmologist Feedback}
In the subjective evaluations, ophthalmologists primarily assessed the clarity and shape of blood vessels, with the majority acknowledging that the 3D CycleGAN-3 model preserved blood vessel clarity and shape effectively, as well as the edges. The next aspect they considered was color code preservation, particularly the representation of the green channel, which is crucial for biological interpretation. Attention was also given to background detail, overall quality, aesthetics, and the correct distribution of colors, a critical factor for the biological accuracy of the images. For example, in scoring Set 2 of Figure~\ref{fig:wref}, some experts preferred the 3D CycleGAN-3 for its accuracy in the green channel, compared to the 3D CycleGAN-GL, which displayed slightly more background vessels but less accuracy.

The UNSB model, however, received criticism for incorrect color code preservation. Set 6 of Figure~\ref{fig:woref} was noted for instances where the 2D CycleGAN and UNSB models ignored edges, and in Set 7 of the same figure, the 2D CycleGAN was criticized for exhibiting too much random color, missing the green staining seen in the reference, and unclear imaging.

The feedback from ophthalmologists suggests that the 3D CycleGAN-3 model not only effectively achieved a stylistic modal transfer but also more importantly preserved the biological content of the medical images, which is vital for clinical interpretation and diagnosis.

\paragraph{Analysis of Hallucinations}

Following feedback from ophthalmological evaluations, we now focus on analyzing how accurately our models avoid introducing hallucinations—false features not actually present in the true anatomy.

For subjectively evaluating the model hallucinations, we focus on two key biological features relevant to retinal imaging: vessel structures in the red channel and immune cell markers, CD4+ T cells, in the green channel. These features are crucial for assessing vascular structure and immune responses within the retina for uveitis, respectively.

\begin{figure}[htb]
\centering
\includegraphics[width=0.95\textwidth]{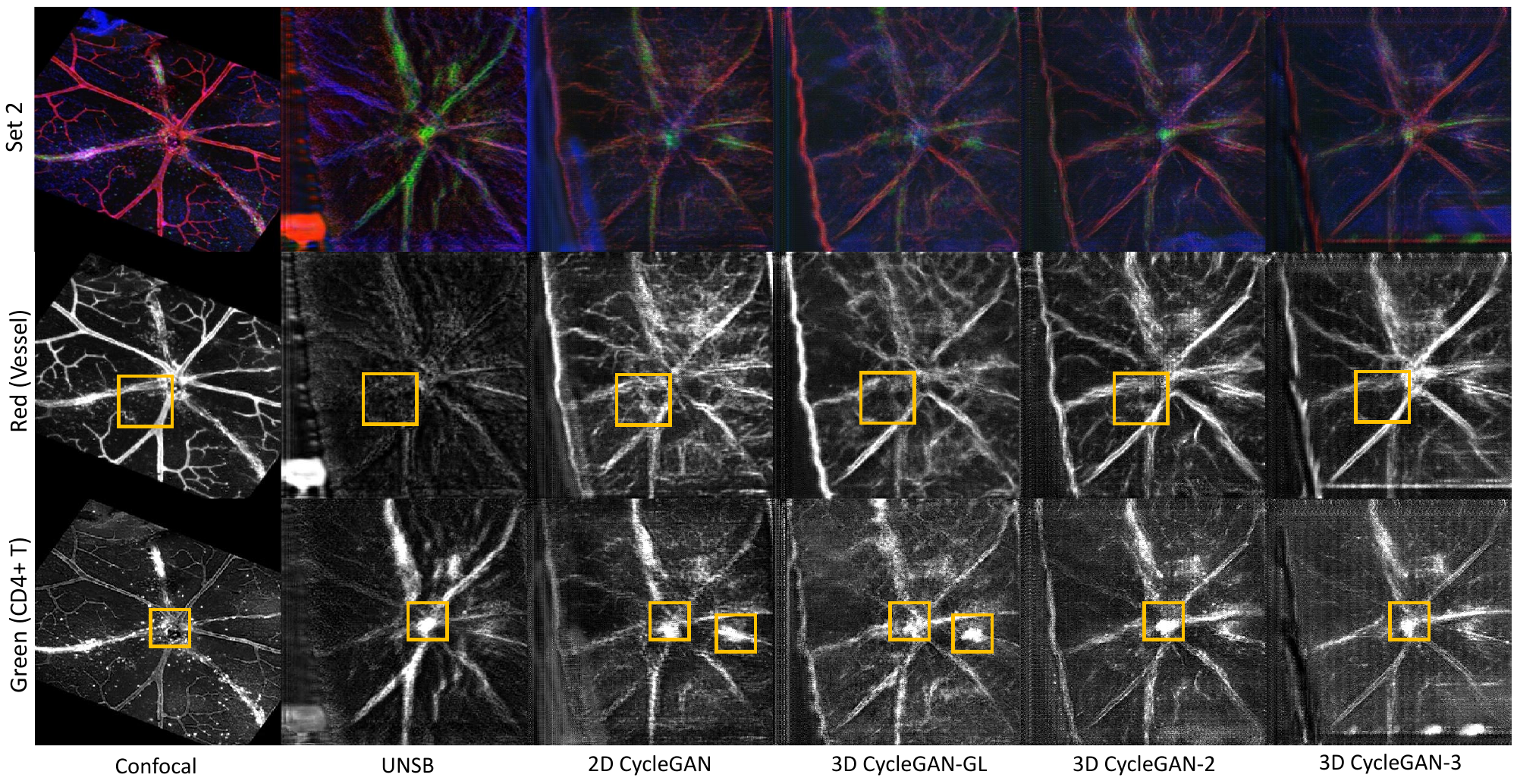}
\caption[Model hallucination analysis.]{Example of model hallucination analysis. Focusing on the red channel for vascular structures and the green channel for CD4+ T cells. Areas highlighted (yellow boxes) show where each model introduces inaccuracies in the representation of vascular and immune cell distributions}
\label{fig:hallucination}
\end{figure}

As shown in Figure~\ref{fig:hallucination}, our analysis in Set 2 reveals notable differences in the clarity and fidelity of vascular structures among the models. The 3D CycleGAN-3 model generally outperforms both the 2D models and other 3D variations in preserving the integrity of major blood vessels with minimal distortions. Specifically, areas highlighted in the red channel exhibit fewer hallucinated vessels, which are incorrectly generated features not aligned with the underlying anatomical structure of the retina.
 
Similarly, in the green channel, which focuses on the distribution of CD4+ T cells indicative of immune activity, the 3D CycleGAN-3 model shows a stronger correlation with the original confocal images in terms of brightness which indicates the immune response areas. However, artifacts around the ONH in the center are present across all models, with our proposed 3D CycleGAN-3 model demonstrating the least severity in artifact generation.

\subsection{Computational Efficiency}

To assess the computational demands of each model, we analyzed the number of parameters (\#Params), number of Floating Point Operations (\#FLOPs), and RunTime (RT) for each translation process. The \#Params and \#FLOPs represent the total number of trainable parameters and floating-point operations required to generate an image, respectively. \#Params measures the model complexity and memory usage. \#FLOPs provides an estimate of computational intensity, crucial for understanding the processing power required and potential latency in real-time applications. The RT is measured during inference, indicating the practical deployment efficiency of each model, reflecting the time taken to process an image. For 2D models, the RT is calculated by summing the times required to process 9 of 512x512 images, simulating the workflow for generating a complete 3D volume from 3D models.

\begin{table}[h]
\centering
\caption{Comparative computational of different models. For each column, the {\color[HTML]{FF0000} red} indicates the most computationally efficient values for each metric.}
\begin{minipage}{\textwidth} 
\centering
\label{tab:computation}
\begin{tabular}{@{}lcccc@{}}
\toprule
Model          & \#Params (M)    & \#FLOPs (G)      & RT (s)         & MOS ↑           \\ \midrule
UNSB           & 14.684          & {\color[HTML]{FF0000}253.829} & 9.891          & 26.269          \\
2D CycleGAN    & {\color[HTML]{FF0000}11.378} & 631.029          & {\color[HTML]{FF0000}0.945} & 40.485          \\
3D CycleGAN-GL & 47.793          & 1323.729         & 32.140         & 49.746          \\
3D CycleGAN-2  & 47.793          & 1323.729         & 35.250         & 53.115          \\
3D CycleGAN-3  & 191.126         & 1585.332         & 94.451         & {\color[HTML]{FF0000}56.469} \\ \bottomrule
\end{tabular}%
\footnotetext{Note: '\#Params' denotes the total number of trainable parameters in millions (M), '\#FLOPs' represents the computational complexity in billions (G) of floating-point operations, and 'RT' indicates the average execution time in seconds (s) per image. Lower values in each metric indicate more efficient computational performance. MOS (Mean Opinion Score) rates the subjective quality of images with higher scores reflecting better quality.}
\vspace{.15in}
\end{minipage}
\end{table}

Table~\ref{tab:computation} illustrates the trade-off between computational efficiency and image quality. While the 2D models (UNSB and 2D CycleGAN) demonstrate quicker processing times, their lower MOS scores suggest a compromise in image quality. In contrast, the extended runtimes associated with 3D models, although potentially limiting for real-time applications, result in higher-quality images that are more clinically valuable, as reflected by their higher MOS scores and positive feedback from ophthalmologists.

The 2D CycleGAN, with the lowest \#Params (11.378M) and moderate \#FLOPs (631.029G) among all models, offers rapid inference times at 0.945s for processing a \( 512 \times 512 \times 9 \) 3D data. This indicates that 2D CycleGAN is more suitable for applications requiring quick image processing such as real-time. However, as revealed by its MOS and the feedback from quality evaluations, the lower complexity cannot adequately capture the spatial relationships and structural complexity inherent in 3D data.

Despite having a higher parameter count than a 2D model, the UNSB model exhibits relatively fewer \#FLOPs (253.829G), which may be attributed to its diffusion-based generative process. Although this process involves numerous iterations, each iteration consists of simpler operations, thus accumulating a lower total computational load (\#FLOPs). However, the need for multiple iterations to refine image quality leads to significantly longer runtimes—up to ten times longer than the 2D CycleGAN—illustrating its inefficiency in time-sensitive scenarios.

The 3D CycleGAN models involve substantially higher \#Params and \#FLOPs. Particularly the 3D CycleGAN-3, with 191.126M \#Params and 1585.332G \#FLOPs and the longest RT of 94.451s. However, this investment in computational resources facilitates a more accurate rendering of complex 3D structures, as evidenced by its highest MOS of 56.469, suggesting superior image quality and detail retention. However, the increased computational demands of 3D models present a challenge for real-time applications, where quick processing is essential. Therefore, future efforts will focus on optimizing the computational efficiency of 3D models without compromising their ability to deliver high-quality 3D image translations to enable time-sensitive applications.

\section{Conclusion and Future Work}
\label{sec:conclusion}
In this paper, we present the 3D CycleGAN framework as an effective tool for translating information inherent in OCT images to the confocal domain, thereby effectively bridging in-vivo and ex-vivo imaging modalities. Although limited by dataset size, our quantitative and qualitative experiments showcased the 3D model's superiority over 2D models in maintaining critical image characteristics, such as blood vessel clarity and color code preservation. Our method demonstrates significant potential in providing non-invasive access to retinal confocal microscopy, which could be revolutionary for observing pathological changes, early disease detection, and studying drug responses in biomedical research. Results from our uveitis dataset could help retinal vein occlusion or retinal inflammation observation, as detailed visualization of inflammatory cell distribution (the color distribution) in the retina can provide insights into the inflammatory processes. While the current translation results require further refinement for clinical application, the potential to identify different immune cell types such as lymphocytes and monocytes and layer changes in high-resoltion translated retinal images could notably enhance the assessment of immune responses and pathologic conditions in retinal diseases like AMD and DR directly from OCT scans. Thus, future efforts will focus on expanding the dataset for more accurate and higher resolution outputs and optimizing the 3D framework for computational efficiency, aiming to advance preclinical study, early disease detection and diagnostics. In line with these enhancements, we intend to explore the integration of 2D projections with sampled 3D data for a 3D reconstruction-based OCT to confocal translation. This approach is designed to maintain the 3D spatial information while reducing computational demands. Further development will include adapting the model for human OCT to confocal translation and applying the translated results for early disease detection and enhanced diagnostic practice.

\FloatBarrier

\begin{appendix}

\section{Appendix}\label{appendixA}

\begin{table*}[htb]
\centering
\caption{The performance of models evaluated by DB metrics FID scores and KID scores, alongside the subjective MOS rating of each individual set. The best result is colored in {\color[HTML]{FF0000} red} and the second best is coloured in {\color[HTML]{0922DB} blue}.}
\begin{minipage}{\textwidth} 
\centering
\label{allsubjective}
\resizebox{\textwidth}{!}{%
\begin{tabular}{ll|ccc|cccccccccc|ccc}
\toprule
& \multicolumn{1}{c|}{} & \multicolumn{3}{c|}{W Ref} & \multicolumn{10}{c|}{W/O Ref} & \multicolumn{3}{c}{Total} \\ \cmidrule(lr){3-5} \cmidrule(lr){6-15} \cmidrule(lr){16-18}  
\multirow{-2}{*}{Model} & \multicolumn{1}{c|}{\multirow{-2}{*}{Metric}} & A2L\_D24 & A2R\_D24 & B3R\_D24 & A2R\_D10 & A2R\_D14 & A2R\_D17 & B3L\_D24 & B3R\_D17 & TX12E2L\_D14 & TX12E2R\_D14 & TX12E3L\_D14 & TX13B1L\_D14 & TX13B3R\_D14 & W Ref & W/O Ref & All \\ \midrule
 & FID768 ↓ & 1.768 & 1.699 & 1.510 & 1.975 & 1.638 & 1.567 & 1.643 & 1.620 & 1.372 & 1.454 & 1.594 & 1.641 & 1.603 & 1.659 & 1.611 & 1.622 \\ 
 & FID2048 ↓ & 335.571 & 336.422 & 267.572 & 434.316 & 318.284 & 324.705 & 317.714 & 271.564 & 318.818 & 274.218 & 312.400 & 243.367 & 201.276 & 313.189 & 301.666 & 304.325 \\ 
 & KID ↓ & 0.663 & 0.643 & 0.485 & 0.968 & 0.645 & 0.707 & 0.750 & 0.585 & 0.715 & 0.537 & 0.685 & 0.516 & 0.439 & 0.597 & 0.655 & 0.641 \\ 
\multirow{-4}{*}{UNSB} & MOS ↑ & 31.700 & 22.900 & 33.300 & 26.100 & 19.400 & 25.800 & 26.600 & 30.000 & 33.900 & 24.200 & 22.200 & 26.300 & 19.100 & 29.300 & 25.360 & 26.269 \\ \midrule
 & FID768 ↓ & 1.668 & 1.509 & 1.465 & 1.553 & 1.364 & 1.640 & 1.415 & 1.349 & 1.490 & 1.250 & 1.437 & 1.564 & 1.140 & 1.547 & 1.420 & 1.449 \\  
 & FID2048 ↓ & 215.451 & 240.731 & 219.724 & 260.986 & 250.141 & 247.595 & 217.812 & 214.397 & 254.606 & 216.557 & 254.916 & 215.287 & 178.178 & 225.302 & 231.048 & 229.722 \\ & KID ↓ & 0.284 & 0.327 & 0.288 & 0.383 & 0.353 & 0.351 & 0.339 & 0.293 & 0.366 & 0.283 & 0.364 & 0.300 & 0.226 & 0.300 & 0.326 & 0.320 \\ 
 \multirow{-4}{*}{2D CycleGAN} & MOS ↑ & 37.000 & 42.300 & 30.700 & 31.400 & 44.200 & 42.700 & 40.400 & 48.000 & 35.900 & 42.200 & 35.400 & 49.000 & 47.100 & 36.667 & 41.630 & 40.485 \\ \midrule
 & FID768 ↓ & 1.633 & 1.128 & 1.081 & 1.287 & 1.247 & 1.283 & 1.075 & 1.089 & 1.304 & 1.283 & 1.162 & 1.077 & 0.892 & 1.281 & 1.170 & 1.195 \\  
 & FID2048 ↓ & 281.033 & 165.763 & 161.590 & {\color[HTML]{0937E1} 173.151} & 185.445 & 193.396 & 166.591 & {\color[HTML]{0937E1} 154.707} & {\color[HTML]{0937E1} 183.646} & 195.985 & 171.409 & {\color[HTML]{0937E1} 152.085} & {\color[HTML]{FF0000} 119.148} & 202.795 & 169.556 & 177.227 \\  
 & KID ↓ & 0.401 & 0.206 & 0.193 & 0.212 & 0.246 & 0.294 & 0.207 & 0.195 & 0.224 & 0.255 & 0.228 & 0.164 & 0.124 & 0.267 & 0.215 & 0.227 \\  
\multirow{-4}{*}{3D CycleGAN-GL} & MOS ↑ & 46.900 & 42.100 & {\color[HTML]{FF0000} 62.200} & {\color[HTML]{FF0000} 63.300} & 39.000 & {\color[HTML]{0922DB} 46.300} & 50.400 & 42.800 & 46.100 & 47.100 & {\color[HTML]{0922DB} 47.400} & 50.800 & {\color[HTML]{0922DB} 62.300} & 50.400 & 49.550 & 49.746 \\ \midrule
 & FID768 ↓ & {\color[HTML]{0922DB} 0.994} & {\color[HTML]{0922DB} 0.840} & {\color[HTML]{FF0000} 0.723} & {\color[HTML]{0922DB} 0.969} & {\color[HTML]{0922DB} 0.927} & {\color[HTML]{0922DB} 0.953} & {\color[HTML]{0922DB} 0.794} & {\color[HTML]{0922DB} 0.793} & {\color[HTML]{0922DB} 0.949} & {\color[HTML]{0922DB} 1.023} & {\color[HTML]{0922DB} 0.859} & {\color[HTML]{0922DB} 0.880} & {\color[HTML]{0922DB} 0.749} & {\color[HTML]{0937E1} 0.852} & {\color[HTML]{0937E1} 0.890} & {\color[HTML]{0937E1} 0.881} \\  
 & FID2048 ↓ & {\color[HTML]{FF0000} 176.433} & {\color[HTML]{0937E1} 140.350} & {\color[HTML]{FF0000} 131.675} & 202.865 & {\color[HTML]{0937E1} 174.571} & {\color[HTML]{0937E1} 165.633} & {\color[HTML]{0937E1} 140.414} & {\color[HTML]{FF0000} 136.360} & 197.414 & {\color[HTML]{0937E1} 187.966} & {\color[HTML]{0937E1} 171.102} & 153.010 & 135.396 & {\color[HTML]{FF0000} 149.486} & {\color[HTML]{0937E1} 166.473} & {\color[HTML]{0937E1} 162.553} \\  
 & KID ↓ & {\color[HTML]{0922DB} 0.190} & {\color[HTML]{0922DB} 0.136} & {\color[HTML]{FF0000} 0.107} & {\color[HTML]{0922DB} 0.198} & {\color[HTML]{FF0000} 0.161} & {\color[HTML]{0922DB} 0.193} & {\color[HTML]{FF0000} 0.121} & {\color[HTML]{FF0000} 0.125} & {\color[HTML]{FF0000} 0.176} & 0.238 & {\color[HTML]{FF0000} 0.159} & {\color[HTML]{FF0000} 0.110} & {\color[HTML]{0922DB} 0.115} & {\color[HTML]{FF0000} 0.144} & {\color[HTML]{0937E1} 0.160} & {\color[HTML]{0937E1} 0.156} \\  
\multirow{-4}{*}{3D CycleGAN-2} & MOS ↑ & {\color[HTML]{FF0000} 56.900} & {\color[HTML]{0922DB} 65.300} & 39.700 & 37.500 & {\color[HTML]{FF0000} 62.900} & 45.700 & {\color[HTML]{0922DB} 55.900} & {\color[HTML]{FF0000} 57.600} & {\color[HTML]{0922DB} 55.400} & {\color[HTML]{0922DB} 48.200} & {\color[HTML]{FF0000} 52.900} & {\color[HTML]{FF0000} 62.300} & 50.200 & {\color[HTML]{0937E1} 53.967} & {\color[HTML]{0937E1} 52.860} & {\color[HTML]{0937E1} 53.115} \\ \midrule
 & FID768 ↓ & {\color[HTML]{FF0000} 0.821} & {\color[HTML]{FF0000} 0.730} & {\color[HTML]{0922DB} 0.747} & {\color[HTML]{FF0000} 0.883} & {\color[HTML]{FF0000} 0.793} & {\color[HTML]{FF0000} 0.763} & {\color[HTML]{FF0000} 0.701} & {\color[HTML]{FF0000} 0.715} & {\color[HTML]{FF0000} 0.822} & {\color[HTML]{FF0000} 0.835} & {\color[HTML]{FF0000} 0.825} & {\color[HTML]{FF0000} 0.775} & {\color[HTML]{FF0000} 0.685} & {\color[HTML]{FF0000} 0.766} & {\color[HTML]{FF0000} 0.780} & {\color[HTML]{FF0000} 0.777} \\  
 & FID2048 ↓ & {\color[HTML]{0937E1} 177.323} & {\color[HTML]{FF0000} 139.419} & {\color[HTML]{0937E1} 147.526} & {\color[HTML]{FF0000} 172.697} & {\color[HTML]{FF0000} 161.605} & {\color[HTML]{FF0000} 134.269} & {\color[HTML]{FF0000} 133.621} & 165.109 & {\color[HTML]{FF0000} 166.120} & {\color[HTML]{FF0000} 169.130} & {\color[HTML]{FF0000} 164.375} & {\color[HTML]{FF0000} 150.130} & {\color[HTML]{0937E1} 134.827} & {\color[HTML]{0937E1} 154.756} & {\color[HTML]{FF0000} 155.188} & {\color[HTML]{FF0000} 155.089} \\  
 & KID ↓ & {\color[HTML]{FF0000} 0.187} & {\color[HTML]{FF0000} 0.127} & {\color[HTML]{0922DB} 0.144} & {\color[HTML]{FF0000} 0.179} & 0.167 & {\color[HTML]{FF0000} 0.122} & {\color[HTML]{0922DB} 0.126} & {\color[HTML]{0922DB} 0.172} & {\color[HTML]{0922DB} 0.180} & {\color[HTML]{FF0000} 0.185} & {\color[HTML]{0922DB} 0.182} & {\color[HTML]{0922DB} 0.136} & {\color[HTML]{FF0000} 0.114} & {\color[HTML]{0937E1} 0.153} & {\color[HTML]{FF0000} 0.156} & {\color[HTML]{FF0000} 0.155} \\  
\multirow{-4}{*}{3D CycleGAN-3} & MOS ↑ & {\color[HTML]{0922DB} 54.500} & {\color[HTML]{FF0000} 65.900} & {\color[HTML]{0922DB} 50.200} & {\color[HTML]{0922DB} 57.700} & {\color[HTML]{0922DB} 56.800} & {\color[HTML]{FF0000} 58.900} & {\color[HTML]{FF0000} 60.000} & {\color[HTML]{0922DB} 50.600} & {\color[HTML]{FF0000} 56.600} & {\color[HTML]{FF0000} 56.100} & 40.300 & {\color[HTML]{0922DB} 55.500} & {\color[HTML]{FF0000} 71.000} & {\color[HTML]{FF0000} 56.867} & {\color[HTML]{FF0000} 56.350} & {\color[HTML]{FF0000} 56.469} \\ 
\bottomrule
\end{tabular}
}
\footnotetext{Note: FID768 and FID2048 refer to the Fréchet Inception Distance computed with 768 and 2048 features, respectively. KID refers to the Kernel Inception Distance. Both FID and KID indicate better performance with lower scores. Mean Opinion Score (MOS) rates the subjective quality of images with higher scores reflecting better quality.}
\vspace{.15in}
\label{tab:eachset}
\end{minipage}
\end{table*}

\end{appendix}

\FloatBarrier

\section*{Ethical Standards}
All mice experiments were approved by the local Animal Welfare and Ethical Review Board (Bristol AWERB), and were conducted under a Home Office Project Licence.

\section*{Acknowledgements}
We would like to thank all the people from Bristol VI-Lab for their positive input and fruitful discussions during this project. We thank researchers from the Autoimmune Inflammation Research (AIR) group at the University of Bristol for their expert support.

\section*{Funding Statement}
Xin Tian was supported by grants from the China Scholarship Council (CSC).

\section*{Competing Interests}
The authors declare no competing interests exist.

\section*{Data Availability Statement}
The dataset will be made available via the University of Bristol Research Data Storage Facility (UoB RDSF) upon acceptance of the work. Code for training and for using the pre-trained model for translation is available on GitHub at \url{https://github.com/xintian-99/OCT2Confocal_3DCycleGAN}.

%
%
\bibliographystyle{splncs04}
\bibliography{main}
\end{document}